\documentclass[aps,prd,nofootinbib,preprint,floatfix,unsortedaddress]{revtex4-1}
\usepackage{setspace}
\usepackage{graphicx}
\usepackage{dcolumn}
\usepackage{multirow}
\usepackage{subfigure}
\usepackage{times,mathptm}
\usepackage{float}
\usepackage{color}
\usepackage{amsmath,amsfonts}
\usepackage{mathptmx}
\usepackage{mathrsfs}
\usepackage{bbm}
\usepackage{bm}
\usepackage{xfrac}

\newcommand{\beq}{\begin{equation}}
\newcommand{\eeq}{\end{equation}} 
\newcommand{\bea}{\begin{eqnarray}}
\newcommand{\eea}{\end{eqnarray}}

\renewcommand{\d}{\delta}

\renewcommand{\l}{\lambda}

\newcommand{\tK}{\widetilde{K}}

\renewcommand{\b}{\beta}

\newcommand{\tr}{\text{Tr}}

\newcommand{\vx}{{\vec{x}}}
\newcommand{\vy}{{\vec{y}}}

\newcommand{\vk}{{\vec{k}}}

\newcommand{\m}{\mu}

\newcommand{\e}{\epsilon}

\renewcommand{\k}{\kappa}

\renewcommand{\th}{\theta}

\newcommand{\oh}{\frac{1}{2}}

\newcommand{\on}{\frac{1}{9}}
\newcommand{\dg}{\dagger}
\newcommand{\non}{\nonumber}

\newcommand{\rf}[1]{(\ref{#1})}
\newcommand{\ra}{\rightarrow}
\newcommand{\pa}{\partial}
\renewcommand{\vec}[1]{\bm #1}
\usepackage{ulem}

\begin{document}

\title{Comparison of complex Langevin and mean field methods applied to effective Polyakov line models}
 
\author{Jeff Greensite}
\affiliation{\singlespacing Physics and Astronomy Department, \\ San Francisco State
University, San Francisco, CA~94132, USA}

%\author{Kurt Langfeld}
%\affiliation{\singlespacing School of Computing \& Mathematics, University of Plymouth, Plymouth, PL4 8AA, UK}
\date{\today}
\vspace{60pt}
\begin{abstract}

\singlespacing
 
       Effective Polyakov line models, derived from SU(3) gauge-matter systems at finite chemical potential, have a
sign problem.  In this article I solve two such models, derived from SU(3) gauge-Higgs and heavy quark theories by the
relative weights method, over a range of chemical potentials where the sign problem is severe.  Two values of the gauge-Higgs coupling are considered, corresponding to a heavier and a lighter scalar particle.  Each model is solved via the complex Langevin method, following the approach of Aarts and James, and also by a mean field technique.  It is shown that where the results of mean field and complex Langevin agree, they agree almost perfectly.  Where the results of the two methods diverge, it is found that the complex Langevin evolution has
a branch cut crossing problem, associated with a logarithm in the action, that was pointed out by M{\o}llgaard and Splittorff.  

\end{abstract}

\maketitle

\singlespacing
%\begin{widetext}
\section{\label{intro}Introduction}

   A recent article by Langfeld and myself \cite{Greensite:2014isa} explains how to extract an effective Polyakov line action from an underlying SU(3) lattice gauge theory, both at zero and at finite chemical potential $\mu$, by the ``relative weights'' method.  The motivation is that the sign problem in the effective theory may be more tractable than the sign problem in the underlying theory, but so far this is only a hope.   In ref.\ 
\cite{Greensite:2014isa} it was also shown how to solve the effective theory via a mean field approach, but mean field results are often unreliable, and therefore the utility of the effective models for finite density investigations is still unknown.  In this article I will consider several effective theories, corresponding to gauge-Higgs and heavy quark models on the lattice, solve those theories at finite chemical potential using both complex Langevin and mean field techniques, and compare the results obtained from each method. 

   The effective Polyakov line action (PLA) corresponding to an underlying lattice gauge theory is the action which results after all degrees of freedom are integrated out, under the constraint that the Polyakov line holonomies are held fixed.  It is convenient to implement this constraint in temporal gauge, which means that the timelike links $U_0(\vx,t)$ on some particular timeslice,
at $t=0$ say, are fixed to the holonomies.  All other timelike links are set to the unit matrix.  If $S_P$ denotes the effective action, $S_L$ the action of the underlying lattice gauge theory, and $\phi$ denotes any matter fields, scalar or fermionic, in the theory, 
then\footnote{Our sign convention for Euclidean actions is chosen so that
the Boltzman weight is proportional to $\exp[+S]$.}
\bea
\exp\Bigl[S_P[U_{\vx},U^\dg_{\vx}]\Bigl] =    \int  DU_0(\vx,0) DU_k  D\phi ~ \left\{\prod_{\vx} \d[U_{\vx}-U_0(\vx,0)]  \right\}
 e^{S_L} \ .
\label{S_P}
\eea
The PLA $S_P$ depends only on the Polyakov line holonomies $U_\vx$, and belongs to the class of SU(3) spin models.
The simplest example of this type of theory, with only nearest-neighbor couplings, is
\bea
          S_{spin} =  J \sum_x \sum_{k=1}^{3}\Bigl(\tr[U_x] \tr[U^\dg_{x+\hat{k}}] + \text{c.c.}\Bigr) +
                                h \sum_x \Bigl( e^{\m/T} \tr[U_x] + e^{-\m/T} \tr[U^\dg_x] \Bigr) \ ,
\label{spin}
\eea
which is in fact the result for an underlying SU(3) gauge theory at finite chemical potential to leading order in a strong coupling/hopping parameter expansion.  Higher orders in this expansion can be found in \cite{Fromm:2011qi}.  At weaker gauge couplings it turns out that each SU(3) spin in the action is coupled to very many spins in its vicinity, and not simply to the nearest neighbors \cite{Greensite:2014isa}.  

    The nearest-neighbor SU(3) spin model has been solved by a number of techniques, including the dual representation 
\cite{Mercado:2012ue}, stochastic quantization \cite{Aarts:2011zn}, reweighting \cite{Fromm:2011qi}, and the mean field approach \cite{Greensite:2012xv}.   For the effective PLA with quasi-local couplings the dual representation method is not applicable, because not all terms in the action have the same sign.  Reweighting, even when supplemented by a cumulant expansion \cite{Ejiri:2013lia}, is also suspect  if the sign problem is really severe \cite{Greensite:2013gya}. This leaves complex Langevin and mean field theory, and it is worth trying out both techniques.

    In this article I will simply write down the effective theories under consideration. How these actions are arrived at via the relative weights method, and the details of the complex Langevin and mean field techniques, may be found in the following references:
\begin{enumerate}
\item {\bf Relative Weights:}  The relative weights method allows one to compute the derivative $dS_P/d\l$ with respective
to some parameter $\l$ which varies the Polyakov line holonomies in the neighborhood of any given field configuration.
By taking derivatives with respect to Fourier components of Polyakov line configurations, it is possible to deduce $S_P$ itself.
The method was developed in a series of articles \cite{Greensite:2013bya,*Greensite:2013yd,*Greensite:2012dy}, and applied to theories with a finite chemical potential in \cite{Greensite:2014isa}.
\item {\bf Complex Langevin:}  The complex Langevin method was applied to the nearest-neighbor SU(3) spin model
by Aarts and James \cite{Aarts:2011zn}.  The effective action $S_P$ can only depend on two linearly independent eigenvalues of each SU(3) holonomy, denoted $e^{i\th_1(\vx)}$ and $e^{i\th_2(\vx)}$, and the angles $\th_{1,2}(\vx)$ are the degrees of freedom which Aarts and James complexify in the Langevin approach
applied to $S_{spin}$.  I follow their method closely, including the use of adaptive step sizes described in 
\cite{Aarts:2009dg},  for solving the more complicated $S_P$ actions considered here.
\item {\bf Mean Field Theory:}  A generalization of the usual mean field approach to the complex action $S_{spin}$  was carried out in ref.\ \cite{Greensite:2012xv} by Splittorff and myself, and the method can be readily applied to more complicated SU(3) spin models with quasi-local couplings, as explained in \cite{Greensite:2014isa}.
\end{enumerate}

    What will be shown is that when the results of mean field field and complex Langevin agree, in the cases considered here, they agree almost perfectly for such observables as Polyakov lines and particle number density.  In the case where the two methods are in strong disagreement, it is found that the complex Langevin approach is invalidated, at the large chemical potential values, by the appearance of a ``branch cut crossing problem'' in Langevin evolution.  This problem refers to the
existence of a branch cut in a logarithm in the action.  If Langevin evolution frequently crosses that branch cut, this can lead to incorrect results for observables, as first noted by M{\o}llgaard and Splittorff  \cite{Mollgaard:2013qra}.

\section{The Models}

   I consider two models at fixed couplings (where the effective PLA has been derived in \cite{Greensite:2014isa}), but variable chemical potential.  The first is the gauge-Higgs system
\bea
     S_L = {\b \over 3} \sum_{p} \text{ReTr}[U(p)] + 
               {\k \over 3} \sum_{x}\sum_{\m=1}^4 \text{Re}\Bigl[\Omega^\dg(x) U_\m(x) \Omega(x+\hat{\m})\Bigr] \ ,
\label{ghiggs}
\eea
at $\b=5.6, ~ \k=3.8$ and $\k=3.9$, and inverse temperature $N_t=6$ lattice spacings in the time direction.  Here $\Omega(x)$ is an SU(3) unimodular scalar field $\Omega^\dg(x) \Omega(x) = 1$, transforming under gauge transformations  $\Omega(x) \ra g(x) \Omega(x)$ in the fundamental representation.   At $\b=5.6$ there is a crossover from ``confinement-like'' to ``Higgs-like'' behavior around $\k=4.0$, where the former type of behavior is similar to QCD:  a linear potential over a finite distance range, followed by string breaking.  In the Higgs-like region there is no linear potential over any interval.  The gauge-Higgs coupling $\kappa=3.9$, since it is closer to the crossover, corresponds to a scalar particle which is
lighter than the scalar particle at $\k=3.8$, although at both couplings the system is in the confinement-like regime. At
$\k=3.8$ the effective PLA was determined to be
\beq
S_P = {1\over 9}\sum_{xy} \tr[U_\vx] \tr[U_\vy^\dg] K(\vx-\vy) + {1\over 3} \sum_x \Bigl\{d_1 e^{\m/T} \tr[U_\vx] +  
         d_1 e^{-\m/T}  \tr[U^\dg_\vx] \Bigr\} \ ,
\label{SP38}
\eeq
where the center symmetry-breaking terms proportional to $d_1$ are identical to those in the SU(3) spin model.  More complicated terms are certainly possible, and may become relevant at sufficiently large values of $\m$, but they are not large enough to be detected at these couplings by the relative weights method, at least with present statistics.
For the lighter particle at $\b=3.9$ an additional term was detected, and the effective action has the form
\bea
S_P &=& {1\over 9}\sum_{xy} \tr[U_\vx] \tr[U_\vy^\dg] K(\vx-\vy)
\non \\
 & &  + {1\over 3}\sum_x \Bigl\{(d_1e^{\m/T}-d_2 e^{-2\m/T}) \tr[U_\vx] +  
          (d_1e^{-\m/T}-d_2 e^{2\m/T}) \tr[U^\dg_\vx] \Bigr\} \ .
\label{SP39}
\eea
On the other hand, the $d_2$ dependent terms originate from double-winding terms
\beq
 {1\over 6} d_2 e^{2\m/T} \tr[U_\vx^2] +  {1\over 6} d_2 e^{-2\m/T} \tr[U_\vx^{\dg 2}]
\eeq
as explained in \cite{Greensite:2014isa}.  Applying the SU(3) identities
\beq
\tr[U_\vx^2] =\tr[U_\vx]^2 -  2\tr[U^\dg_\vx]  ~~~,~~~ \tr[U_\vx^{\dg 2}] =\tr[U^{\dg}_\vx]^2 -  2\tr[U_\vx] \ ,
\label{identities}
\eeq
and neglecting the terms quadratic in $\tr[U],\tr[U^\dg]$, gives \rf{SP39}.  Although neglecting the quadratic terms works nicely at $\m=0$, in the sense that that Polyakov line correlators computed in the effective theory agree with those in the underlying gauge-Higgs theory, it leads to the unphysical result that particle density is increasingly negative with increasingly positive $\m$, as we will see below.  Therefore we also consider the action which we would have without discarding the quadratic terms,
namely
\bea
S_P &=& {1\over 9}\sum_{xy} \tr[U_\vx] \tr[U_\vy^\dg] K(\vx-\vy) + {1\over 3} \sum_x \Bigl\{d_1 e^{\m/T} \tr[U_\vx] +  
         d_1 e^{-\m/T}  \tr[U^\dg_\vx] \Bigr\} 
\non \\
& & + {1\over 6} \sum_x \Bigl\{d_2 e^{2\m/T} \tr[U_\vx^2] +  
         d_2 e^{-2\m/T}  \tr[U^{2 \dg}_\vx] \Bigr\} 
\label{SP39q}
\eea

  The quasi-local kernel $K(\vx-\vy)$ is given by the form
\beq
    K(\vx-\vy) = \left\{ \begin{array}{cl}
                   {1\over L^3}\sum_\vk \tK^{fit}(k_L) e^{-i\vk\cdot (\vx-\vy)} & |\vx-\vy| \le r_{max} \cr \\
                      0 & |\vx-\vy| > r_{max} \end{array} \right. \ ,
\label{K1}
\eeq
where
\bea
          \tK^{fit}(k_L) = \left\{ \begin{array}{cl}
                 \oh c_1 - 2 c_2 k_L & k_L \le k_0 \cr \\
                 \oh b_1 - 2 b_2 k_L & k_L > k_0 \end{array} \right. \ ,
\label{K2}
\eea
and $k_L$ is the magnitude of lattice momentum. 
\beq
           k_L = 2 \sqrt{ \sum_{i=1}^3 \sin^2(k_i/2)} \ .
\eeq 
Components $k_i$ are wavenumbers on the three-dimensional lattice.
At couplings $\b=5.6$ and $\k=3.8,3.9$ the various parameters which define the effective model are given in Table \ref{tab1}. 

\begin{table}[t!]
\begin{center}
\begin{tabular}{|c|c|c|c|c|c|c|c|c|c|c|} \hline
          model  & $\k$ &  $c_1$ & $c_2$ & $k_0$ &  $b_1$ & $b_2$ & $r_{max}$ & $d_1$ & $d_2$ & $h$ \\
\hline
        gauge-Higgs & 3.8 &  9.77(8) & 1.18(2)  & 1.63 & 6.77(17) & 0.72(2)   & $\sqrt{41}$ & 0.0195(4) & $< 0.001$ & NA \\ 
        gauge-Higgs & 3.9 &  12.55(13) & 1.69(4)  & 1.36 & 8.16(17) & 0.89(2)   & no cutoff & 0.0585(8) & 0.0115(2) & NA \\ 
        heavy quark & NA &  7.15(5) &  0.79(1) & 1.79 & 6.22(14) & 0.66(1)  & $\sqrt{29}$ &  NA &NA &  $10^{-4}$  \\
\hline
\end{tabular}
\caption{Parameters defining the effective Polyakov line action $S_P$ corresponding to (i) the SU(3) gauge-Higgs theory at 
$\b=5.6$ on a $16^3 \times 6$ lattice with $\k=3.8,3.9$ and (ii) the heavy quark model at $\b=5.6$ on a  $16^3 \times 6$ lattice.}
\label{tab1}
\end{center}
\end{table}
 
  The second model is the heavy quark model.    Let $\zeta$ represent the hopping parameter for Wilson fermions, or $1/2m$ for staggered fermions, and $h=\zeta^{N_t}$.  In the limit that $\zeta \ra 0$ and $e^\mu \ra \infty$ in such a way that $\zeta e^\mu $ is finite, the lattice action simplifies drastically \cite{Bender:1992gn,*Blum:1995cb,*Engels:1999tz,*DePietri:2007ak}.  In temporal gauge,
\bea
\exp[S_L] = \prod_\vx \det\Bigl[1+h e^{\m/T} U_0(\vx,0)\Bigr]^p \det\Bigl[1+h e^{-\m/T} U^\dg(\vx,0) \Bigr]^p \exp[S_{plaq}] \ ,
\eea
where $p=1$ for four-flavor staggered fermions, $p=2N_f$ for Wilson fermions ($N_f$ is the number of flavors), and where
the determinant refers to color indices since the Dirac indices have already been accounted for.  $S_{plaq}$ is the usual Wilson action of the pure gauge theory.  The corresponding PLA $S_P$ is given by
\bea
\exp[S_P] = \prod_\vx \det\Bigl[1+h e^{\m/T} U_\vx \Bigr]^p \det\Bigl[1+h e^{-\m/T} U^\dg_\vx \Bigr]^p   \exp[S^0_P] \ ,
\eea
where determinants can be expressed entirely in terms of Polyakov line operators, using the identities
\bea
\det\Bigl[1+h e^{\m/T} U_\vx \Bigr] &=& 1 + he^{\m/T} \tr[U_\vx] + h^2 e^{2\m/T} \tr[U_\vx^\dg] + h^3 e^{3\m/T} \ ,
\non \\
\det\Bigl[1+h e^{-\m/T} U^\dg_\vx \Bigr] &=& 1 + he^{-\m/T} \tr[U_\vx^\dg] + h^2 e^{-2\m/T} \tr[U_\vx] + h^3 e^{-3\m/T} \ ,
\eea
and $S^0_P$ is the effective action of the pure lattice gauge theory at the given $\b$
\beq
S^0_P = {1\over 9} \sum_{xy} \tr[U_\vx] \tr[U_\vy^\dg] K(\vx-\vy) \ ,
\label{S0}
\eeq
with $K(\vx-\vy)$ defined by eqs.\ \rf{K1} and \rf{K2}.  We work with four staggered fermions ($p=1$) at $\b=5.6$ and $h=10^{-4}$, with inverse temperature $N_t=6$.  The constants needed to compute the kernel $K(\vx-\vy)$ in this case are given in the third row of Table \ref{tab1}.  Bringing the determinants into the action, we have
\bea
S_P &=& {1\over 9} \sum_{xy} \tr[U_\vx] \tr[U_\vy^\dg] K(\vx-\vy) + 
\sum_\vx \left\{ \log\big(1 + he^{\m/T} \tr[U_\vx] + h^2 e^{2\m/T} \tr[U_\vx^\dg] + h^3 e^{3\m/T}\big) \right.
\non \\
& & \qquad \left. + 
\log\big(1 + he^{-\m/T} \tr[U_\vx^\dg] + h^2 e^{-2\m/T} \tr[U_\vx] + h^3 e^{-3\m/T}\big) \right\} \ .
\eea

\section{The Methods}

\subsection{The complex Langevin approach}
    The PLA $S_P$ inherits, from the underlying gauge theory, an invariance under local transformations
\beq
             U_\vx \ra  g_\vx U_\vx g^\dg_\vx \ ,
\label{symmetry}
\eeq
where $g_\vx$ is a position-dependent element of the gauge group.  This means that $S_P$ can depend on holonomies
only through local traces of powers of holonomies $\tr[U^p_\vx]$; there can be no dependence on expressions such as 
$\tr[U_\vx U_\vy]$, since for $\vx \ne \vy$ this term is not invariant under \rf{symmetry}.  Equivalently, the invariance \rf{symmetry} means that $S_P$ depends only on the eigenvalues $e^{i\th_a(\vx)}$ of the holonomies $U_\vx$. In particular,
\beq
          \tr[U_\vx] = e^{i\th_1(\vx)} + e^{i\th_2(\vx)} + e^{-i(\th_1(\vx) + \th_2(\vx))} \ .
\eeq
In the complex Langevin approach  \cite{Aarts:2011zn} the angles $\{\th_a(\vx),a=1,2\}$ are treated as the dynamical variables, which means that, for purposes of stochastic quantization, the Haar measure 
\beq
dU = d\th_1 d\th_2 \sin^2\left({\th_1-\th_2 \over 2}\right)\sin^2\left({2\th_1+\th_2 \over 2}\right)\sin^2\left({\th_1+2\th_2 \over 2}\right)
\eeq
must be incorporated into the action of the effective PLA, i.e.
\beq
S_P \longrightarrow S'_P = S_P + \sum_\vx \log  \left\{  \sin^2\left({\th_1(\vx)-\th_2(\vx) \over 2}\right)  \sin^2\left({2\th_1(\vx)+\th_2(\vx) \over 2}\right)  \sin^2\left({\th_1(\vx)+2\th_2(\vx) \over 2}\right) \right\} \ .
\label{SP'}
\eeq
The prescription is then to complexify the angles,
\beq
         \th_a(\vx) = \th^R_a(\vx) + i \th^I_a(\vx) ~~,~~ a=1,2 \,
\eeq
and solve the complex Langevin equation, which is a first-order differential equation in the fictitious Langevin time $t$.
Discretizing the Langevin time, $t_n = n \e$, the complex Langevin equation is\footnote{Note that the unconventional 
plus sign in front of $\pa S/\pa \th$ follows from our unconventional sign convention for the action (see footnote 1).}
\bea
       \th^R_a(\vx,t_{n+1}) &=& \th^R_a(\vx,t_n) + \mbox{Re}\left\{ \left({\pa S'_P[\th,t_n] 
              \over \pa \th_a(\vx,t_n)}\right)_{\th\ra\th^R+i\th^I} \right\} \e + \eta_a(\vx,t_n) \sqrt{\e} \ ,
\non \\
       \th^I_a(\vx,t_{n+1}) &=& \th^I_a(\vx,t_n) + \mbox{Im}\left\{ \left({\pa S'_P[\th,t_n] 
              \over \pa \th_a(\vx,t_n)}\right)_{\th\ra\th^R+i\th^I} \right\}  \e \ ,
\label{cle}
\eea
where $\eta_a(\vx,t_n)$ is a (real-valued) random variable satisfying
\beq
     \langle \eta_a(\vx,t_n) \eta_b(\vy,t_m) \rangle =  2 \d_{\vx \vy}\d_{nm} \d_{ab} \ .
\eeq
In solving this equation it is important to use an adaptive stepsize in order to prevent runaway solutions, 
reducing $\e$ when the magnitude of  $\pa S_P/ \pa \th$ becomes large at any lattice site, as explained in 
\cite{Aarts:2009dg}.\footnote{Aarts and James \cite{Aarts:2011zn} also implemented an improved version of the Langevin equation, to reduce the dependence on stepsize $\e$.  The results reported in the next section were obtained using the unimproved version \rf{cle} of the complex Langevin equation (with an adaptive stepsize).}

   There is a danger in applying the complex Langevin equation to actions which incorporate a logarithm, as already mentioned. The problem has been pointed out by 
M{\o}llgaard and Splittorff \cite{Mollgaard:2013qra}.  Logarithms have a branch cut along the negative real axis in the complex plane, and if, after complexification of the field variables, the argument of the logarithm repeatedly crosses the negative real axis in the course of Langevin evolution, then the results for observables are unreliable.  The effective action corresponding to gauge-Higgs theory contains a logarithm of the Haar measure, and the effective action for the heavy quark model contains, in addition, a logarithm of the fermion determinant.  The Langevin evolution of the argument of these logarithms in the complex plane must therefore be monitored, to check that the crossings of the branch cut are negligible.

\subsection{\label{mf}Mean field theory}

     In mean field theory, the basic idea is to localize the part of the action which depends on products of SU(3) spins at different sites. For the effective actions we consider here, these products are contained in the quasilocal term
\bea
           S^0_P &=& {1\over 9} \sum_{\vx \vy} \tr[U_\vx] \tr[U_\vy^\dg] K(\vx-\vy)
\non \\
                       &=&  {1\over 9} \sum_{(\vx \vy)} \tr[U_\vx] \tr[U_\vy^\dg] K(\vx-\vy) + 
                               a_0 \sum_{\vx} \tr[U_\vx] \tr[U_\vx^\dg] \ ,
\eea
where we have introduced the notation for the double sum, excluding $\vx=\vy$,
\beq
    \sum_{(\vx \vy)} \equiv \sum_{\vx} \sum_{\vy \ne \vx}  ~~~~\text{and}~~~~ a_0 \equiv {1\over 9} K(0) \ .
\eeq
Next, following the treatment in ref.\ \cite{Greensite:2012xv}, we introduce parameters $u,v$
\beq
   \tr U_\vx = (\tr U_\vx - u) + u  ~~~,~~~ \tr U^\dg_\vx = (\tr U^\dg_\vx - v) + v 
\eeq
so that
\bea
S_P^0 &=& J_0 \sum_\vx (v \tr U_\vx + u \tr U_\vx^\dg) - uvJ_0V + a_0 \sum_{\vx} \tr[U_\vx] \tr[U_\vx^\dg]+E_0 \ ,
\eea
where $V=L^3$ is the lattice volume, and we have defined
\bea
     E_0 &=& \sum_{(\vx \vy)} (\tr U_x-u)(\tr U_\vy^\dg - v) \on K(\vx-\vy) \ ,
\non \\
     J_0 &=&  {1\over 9} \sum_{\vx \ne 0} K(\vx) \ .
\eea
The trick is to choose $u$ and $v$ such that $E_0$ can be treated as a perturbation, to be ignored as a first approximation.
In particular, $\langle E_0 \rangle = 0$ when
\bea
u = \langle \tr U_x \rangle ~~~,~~~ v = \langle \tr U^\dg_x \rangle \ .
\label{consistency}
\eea
These conditions turn out to be equivalent to stationarity of the mean field free energy with respect to variations in $u,v$.
After dropping $E_0$, the action is local and and the group integrations can be carried out analytically.  This means that
$ \langle \tr U_x \rangle$ and $\langle \tr U^\dg_x \rangle$ are calculable functions of $u,v$, and the conditions \rf{consistency}
are then solved numerically.  In the case that there is more than one solution, the solution with the minimum free energy is chosen.

    Let us first consider the effective action for gauge-Higgs theory \rf{SP39q}.  After discarding the $E_0$ term, the 
action becomes
\bea
S_P &=& J_0 \sum_\vx (v \tr U_\vx + u \tr U_\vx^\dg) - uvJ_0V +  a_0 \sum_{\vx} \tr[U_\vx] \tr[U_\vx^\dg]
 +  {1\over 3}\sum_x \Bigl\{d_1 e^{\m/T} \tr[U_\vx] + d_1 e^{-\m T} tr[U^{\dg}_\vx] \Bigr\} 
\non \\
 & & + \sum_x \Bigl\{a_2 e^{2\m/T} \tr[U_\vx^2] +  
         a_2 e^{-2\m/T}  \tr[U^{2 \dg}_\vx] \Bigr\} 
\non \\
&=& \sum_\vx (A \tr U_\vx + B \tr U_\vx^\dg) - uvJ_0V +  a_0 \sum_{\vx} \tr[U_\vx] \tr[U_\vx^\dg]
 + \sum_x \Bigl\{a_2 e^{2\m/T} \tr[U_\vx^2] + a_2 e^{-2\m/T}  \tr[U^{2 \dg}_\vx] \Bigr\} 
\non \\
\eea
where $a_2=d_2/6$, and we have defined
\beq
A = J_0 v + {1\over 3} d_1 e^{\m/T} ~~~,  ~~~ B = J_0 u + {1\over 3} d_1 e^{-\m/T} 
\eeq
Denote the mean field partition function (i.e.\ the partition function obtained after dropping $E_0$) as $Z_{mf}=\exp[-f_{mf}V/T]$.
Then 
\bea
Z_{mf} &=& e^{-uvJ_0 V} \bigg\{ \exp \left[a_0 {\pa^2 \over \pa A \pa B}\right] \int DU \exp\left[ A \tr U + B \tr U^\dg \right.
\non \\
  & & \left. + a_2 e^{2\m/T} \tr U^2 + a_2 e^{-2\m/T}  \tr U^{\dg 2}\right]  \bigg\}^V
\eea
and make the rescalings
\beq
    u = u' e^{-\m/T} ~~,~~ v = v' e^{\m/T} ~~,~~ A = A' e^{\m/T} ~~,~~ B = B' e^{-\m/T} \ ,
\label{rescale}
\eeq
so that
\bea
Z_{mf} &=& e^{-u'v' J_0 V} \bigg\{ \exp \left[a_0 {\pa^2 \over \pa A' \pa B'}\right] I[A',B',a_2] \bigg\}^V
\eea
where
\beq
I[A',B',a_2] =  \int DU \exp\left[ A' e^{\m/T} \tr U + B'  e^{-\m/T} \tr U^\dg + a_2 e^{2\m/T} \tr U^2 + a_2 e^{-2\m/T}  \tr U^{\dg 2}\right] 
\eeq
Repeating the steps in \cite{Greensite:2012xv}, which will not be reproduced here, the SU(3) group integration can be
carried out, and the result for the mean field free energy is
\bea
{f_{mf}\over T} = u'v' J_0  - \log F[A',B',a_2] \ ,
\eea
where 
\bea
F[A',B',a_2] &=& \exp\left[a_0 {\pa^2 \over \pa A' \pa B'} \right]  \sum_{s=-\infty}^{\infty}  e^{3s\m}
\det\Bigl[D^{-s}_{ij} Q(A',B',a_2) \Big] \ ,
\label{F}
\eea
Here $D^{-s}_{ij}$ is the $i,j$-th component of a matrix of differential operators
\bea
D^s_{ij} &=& \left\{ \begin{array}{cl}
                         D_{i,j+s} & s \ge 0 \cr
                         D_{i+|s|,j} & s < 0 \end{array} \right. \ ,
\non \\
D_{ij} &=& \left\{ \begin{array}{cl}
                         \left({\pa \over \pa B'} \right)^{i-j} & i \ge j \cr 
                        \left({\pa \over \pa A'} \right)^{j-i} & i < j \end{array} \right. \ .
\eea
and
\beq
Q[A',B',a_2] = \int {d\phi \over 2\pi} \exp\big[ A' e^{\m/T} e^{i\phi} + B'  e^{-\m/T} e^{-i\phi} +
        a_2 e^{2\m/T} e^{2i\phi} + a_2  e^{-2\m/T} e^{-2i\phi} \big]
\label{angint}
\eeq
When $a_2=0$ the integral can be done exactly, and gives $Q[A',B']=I_0[2\sqrt{A'B'}]$.  For $a_2 \ne 0$ the integration
can be carried out by expanding the integrand in a power series in $A',B',a_2$, with the result
\beq
Q[A',B',a_2] = \sum_{n=0}^\infty \sum_{m=0}^\infty \sum_{l=0}^\infty \sum_{k=0}^\infty {A'^n B'^m a_2^{l+k} \over
n! m! l! k!} \d_{n-m+2l-2k}
\eeq
The mean field free energy is stationary with respect to $u,v$ if the following conditions are
satisfied: 
\bea
u' - {1\over F} {\pa F \over \pa A'} &=& 0 \ ,
\non \\
v' - {1\over F} {\pa F \over \pa B'} &=& 0 \ .
\label{gh-conditions}
\eea
These equations also guarantee the self-consistency conditions \rf{consistency}, and can be solved numerically for 
$u,v$. We then have the mean field results for $\langle \tr[U_\vx] \rangle=u$ and $\langle \tr[U^\dg_\vx] \rangle=v$.
The number density can also be computed from the derivative of the mean field free energy with respect to chemical
potential
\beq
             n = -{\pa f_{mf}\over \pa \mu} \ .
\eeq
 
   There are five infinite sums in the expression for the free energy, with indices denoted $m,n,l,k,s$, which are reduced
to four sums by the Kronecker delta  $\d_{n-m+2l-2k}$.  These sums must be truncated for the numerical evaluation, and then one has to check that the final answers are insensitive to an increase in the cutoff.  In practice one finds that summing
the over indices $n,m,j,k$ from 0 to 12, and cutting off the sum over $s$ at $|s|=6$, is sufficient.  It is also necessary to
expand the operator
\beq
          \exp \left[a_0 {\pa^2 \over \pa A \pa B}\right]
\eeq
in a Taylor series  and truncate the series (third order in $a_0$ is sufficient).

   For the action \rf{SP38} we set $a_2=0$, and use the previous expressions. Although the angular integration 
\rf{angint} has, in this case, the compact result $Q[A',B']=I_0[2\sqrt{A'B'}]$, in practice the computation of \rf{gh-conditions}  is faster using the power series expansion of $Q$.  For the action \rf{SP39}, we define
\beq
A = J_0 v + {1\over 3} \big(d_1 e^{\m/T} - d_2 e^{-2\m/T}\big)  ~~~,  ~~~ 
B = J_0 u + {1\over 3} \big(d_1 e^{-\m/T} - d_2 e^{2\m/T}\big)
\eeq
and refrain from rescaling.  In that case the previous formulas apply with unprimed variables, apart from a modification
\bea
F[A,B] &=& \exp\left[a_0 {\pa^2 \over \pa A \pa B} \right]  \sum_{s=-\infty}^{\infty}  
\det\Bigl[D^{-s}_{ij} Q(A,B,a_2) \Big] \ .
\eea
since the factor $e^{3s\m}$ in \rf{F} comes from the rescaling.

    Finally, in the case of the heavy quark model we define $A=J_0 v, ~ B=J_0 u$, and carry out the rescalings \rf{rescale}.
Then a very similar analysis leads to the conditions        
\bea
u' - {1\over G}{\pa G \over \pa A'}=0  ~~~~~\text{and}~~~~~ v'- {1\over G}{\pa G \over \pa B'}=0 \ ,
\label{hq-conditions}
\eea
where
\bea
G(A',B') &=& \left(a_1 + a_2 e^{-\m/T} {\pa \over \pa A'} + a_3 e^{\m/T} {\pa \over \pa B'} +   
 + a_4 e^{-2\m/T} {\pa^2 \over \pa A'^2} \right.
\non \\
& & \left. + a_5 e^{2\m/T} {\pa^2 \over \pa B'^2}  + a_6 {\pa^2 \over \pa A' \pa B'}\right)^p
 \sum_{s=-\infty}^{\infty} e^{3 \mu s} \det\Bigl[D^{-s}_{ij} I_0[2\sqrt{A' B'}] \Big] \ ,
\eea
and
\bea
a_1 &=& 1 + h^3 (e^{3\mu/T}+e^{-3\mu/T}) + h^6
\non \\
a_2 &=&  (h+h^5)e^{\m/T} + (h^2+h^4)e^{-2\m/T} ~~~,~~~ a_3 =  (h+h^5)e^{-\m/T} + (h^2+h^4)e^{2\m/T} 
\non \\
a_4 &=& h^3 e^{-\m/T} ~~~,~~~ a_5 = h^3 e^{\m/T} ~~~,~~~ a_6 = h^2 + h^3 \ .
\eea
Once again, the mean field conditions \rf{hq-conditions} can be solved numerically, and from the solution we can 
calculate the VEV of the Polyakov lines and the number density as a function of chemical potential.
                             
\section{Results}

\subsection{Gauge-Higgs at $\k=3.8$}

\begin{figure}[htb]
\subfigure[~$\langle \tr(U) \rangle$]  % caption for subfigure a
{   
 \label{u38}
 \includegraphics[scale=0.6]{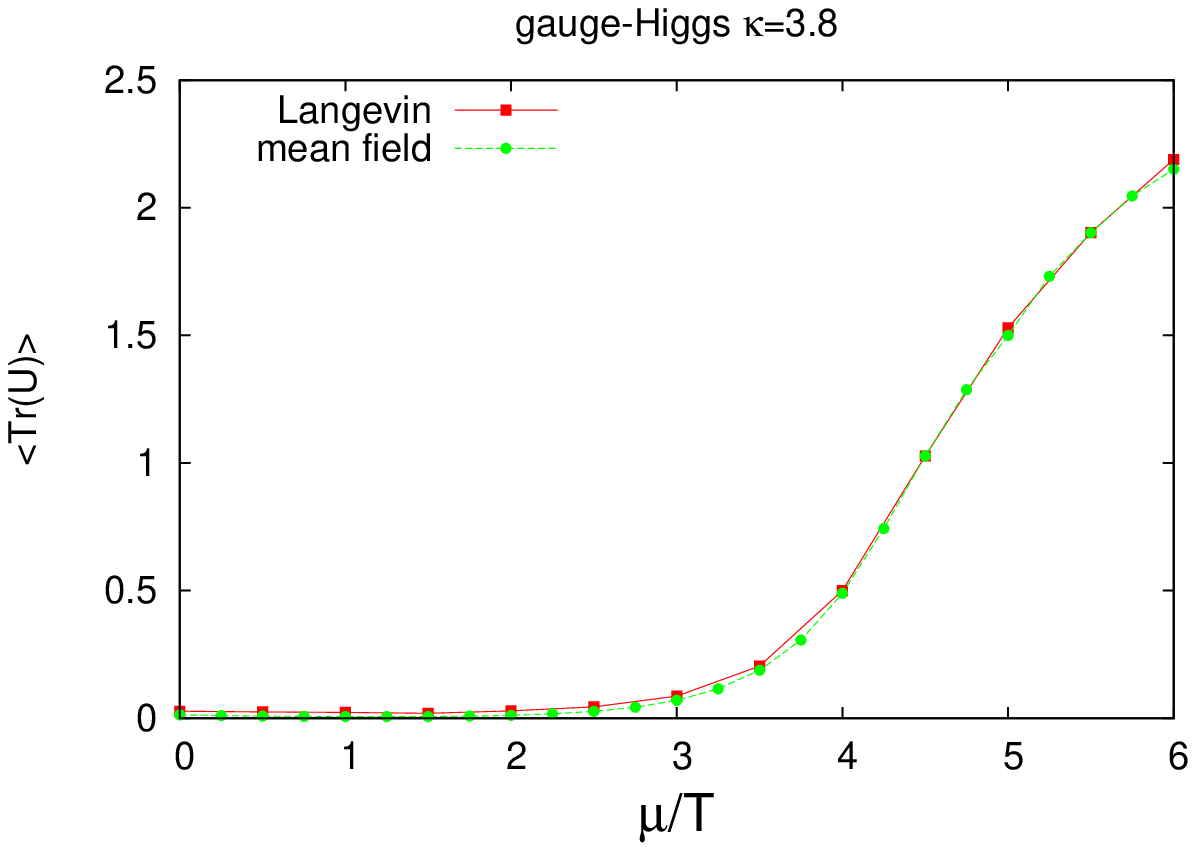}
}
\subfigure[~$\langle \tr(U^\dg) \rangle$]  % caption for subfigure a
{   
 \label{v38}
 \includegraphics[scale=0.6]{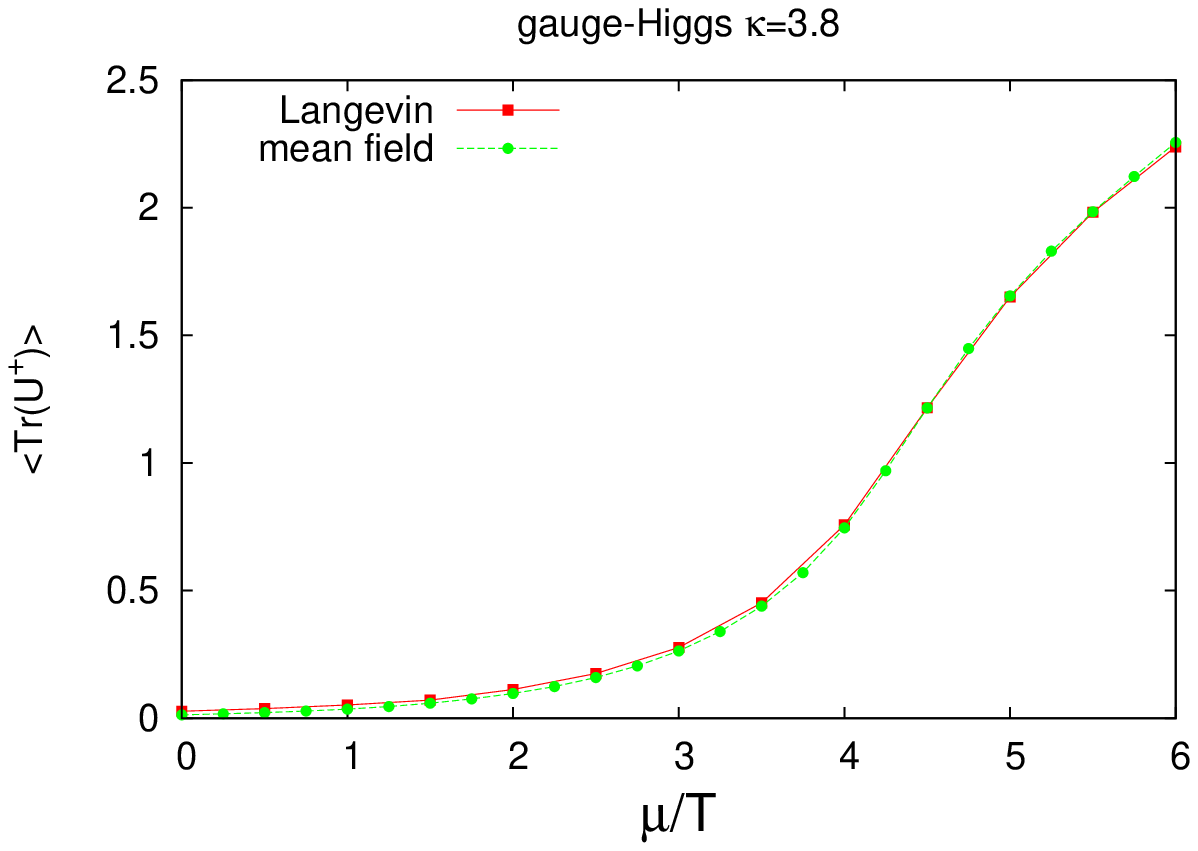}
}
\subfigure[~density]  % caption for subfigure a
{   
 \label{n38}
 \includegraphics[scale=0.6]{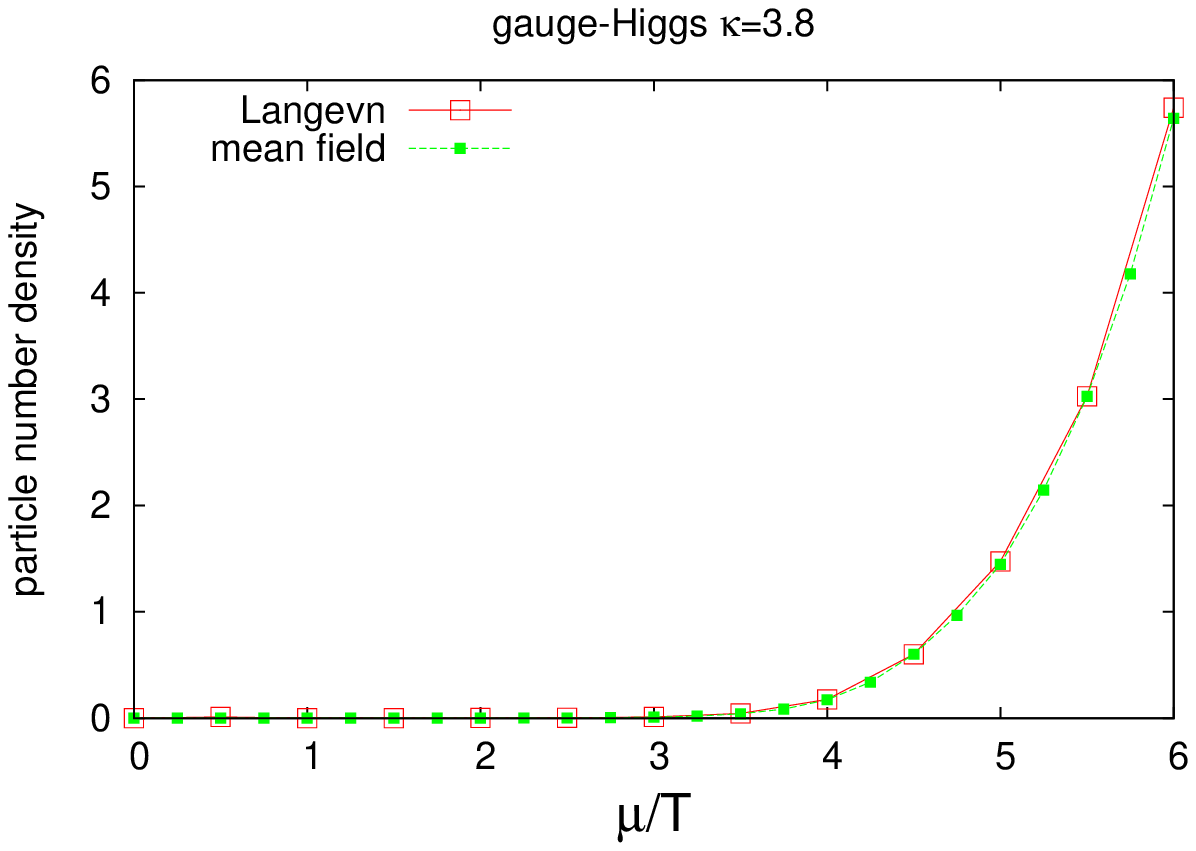}
}
\caption{Comparison of Polyakov lines $\langle \tr(U) \rangle, \langle \tr(U^\dg) \rangle$ and number density vs.\ $\m/T$, computed via  complex Langevin and mean field techniques, in gauge-Higgs theory at $\kappa=3.8$.}
\label{gh38}
\end{figure}

    We begin with the gauge-Higgs model at $\b=5.6, \k=3.8$, and inverse temperature $N_t=6$ lattice spacings; in this
case $d_2=0$ in the effective actions.  Figures
\ref{u38}-\ref{n38} compare the results for Polyakov line expectation values $\langle \tr[U_\vx] \rangle, \langle \tr[U^\dg_\vx] \rangle$ and number density obtained from the complex Langevin equation and from mean field theory.
The numerical agreement is such that the data points derived by each method can barely be distinguished from one another.

\begin{figure}[htb]
\centerline{\scalebox{0.6}{\includegraphics{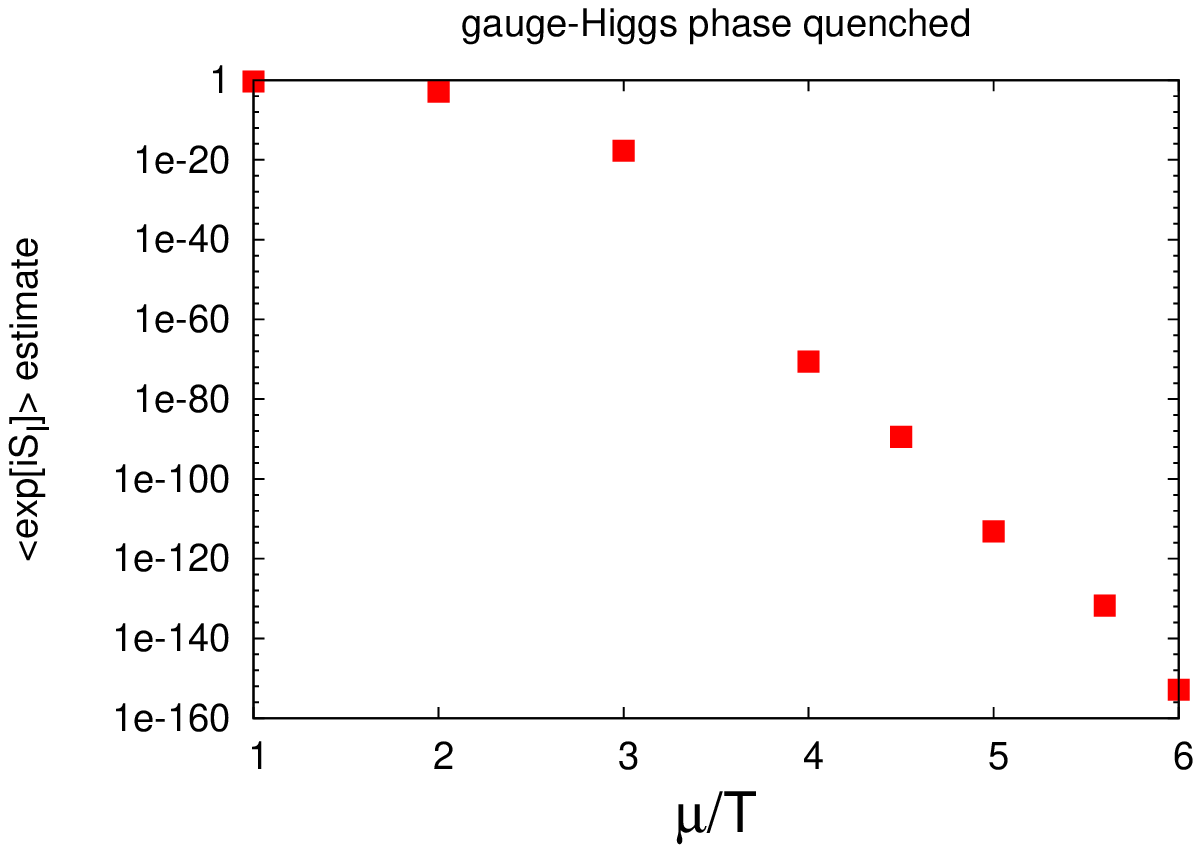}}}
\caption{An estimate of $\langle \exp[iS_I] \rangle_{pq}$ vs. $\m/T$ in the phase-quenched version of gauge-Higgs theory at
$\k=3.8$, obtained from the second order cumulant.  $S_I$ is the imaginary part of the action.}
\label{phase38}
\end{figure}

    An estimate of the severity of the sign problem is provided by a measurement of $\langle e^{iS_I} \rangle_{pq}$, where
$S_I$ is the imaginary part of the action, and the expectation value is taken in the ``phase-quenched'' probability measure 
proportional (in our sign convention) to $e^{S_R}$, where $S_R$ is the real part of the action.  When the sign problem is severe, the expectation value
of $\langle e^{iS_I} \rangle_{pq}$ is so small that it is difficult to distinguish statistically from zero.   However, according to the
cumulant expansion
\beq
             \langle \exp[iS_I] \rangle_{pq} = \exp\left[-\sum_{n=1}^\infty {C_{2n}\over (2n)!}\right] \,
\label{cumulant}
\eeq
where $C_{2n}$ is the $2n$-th order cumulant.  We can therefore get a very rough estimate of the severity of the sign problem
just by truncating the expansion at the second order cumulant $C_{2}=\langle S_I^2 \rangle_{pq}$.   There is, of course, no guarantee that higher cumulants are negligible compared to the second order cumulant, so this truncation may not be very accurate for the observable \rf{cumulant}; perhaps it is within a factor of two or so in the logarithm of the observable.  That is enough to 
judge the severity of the sign problem as $\m$ increases.  The result for the $16^3$ spatial volume is shown in 
Fig.\ \ref{phase38}.

\begin{figure}[htb]
\label{det38m5}
 \includegraphics[scale=0.6]{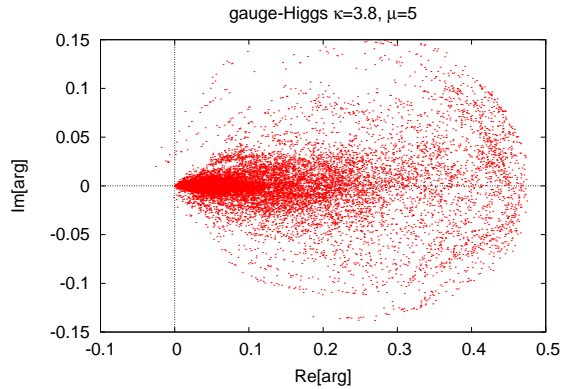}
 \caption{Argument of the logarithm for gauge-Higgs theory at $\b=5.6,~\k=3.8$, at chemical potential
$\m/T=5.0$, evaluated at each Langevin time step.  Values near the negative real axis are a
negligible fraction of the sample.}
\label{detgh38}
\end{figure}

    As mentioned in the previous section, the action which is used in the complex Langevin equation contains the logarithm
of the measure factor (see \rf{SP'}), and it is necessary to monitor this factor to ensure that it only rarely crosses the negative real
axis.  Of course, at real values of $\th_{1,2}$ this measure factor is strictly positive, but that can change when $\th_{1,2}$ are complexified.  Since the measure factor is in this case a product of measure factors at each lattice site,
we pick an arbitrary site $\vx'$ and record the value of the measure factor
\beq
\mbox{Arg} = \sin^2\left({\th_1(\vx')-\th_2(\vx') \over 2}\right)\sin^2\left({2\th_1(\vx')+\th_2(\vx') \over 2}\right)\sin^2\left({\th_1(\vx')+2\th_2(\vx') \over 2}\right) \ ,
\label{measure}
\eeq
which is the argument of one of the logarithms in the action \rf{SP'}, at each Langevin time.  The result, for $\m=5$ is shown in Fig.\ \ref{detgh38}.  We see that the measure is very strongly concentrated on the positive side of the real axis, which suggests that crossings of the logarithm branch cut are relatively rare events.

\subsection{Heavy quark model}

\begin{figure}[htb]
\subfigure[~$\langle \tr(U)\rangle$]  % caption for subfigure a
{   
 \label{uhq}
 \includegraphics[scale=0.6]{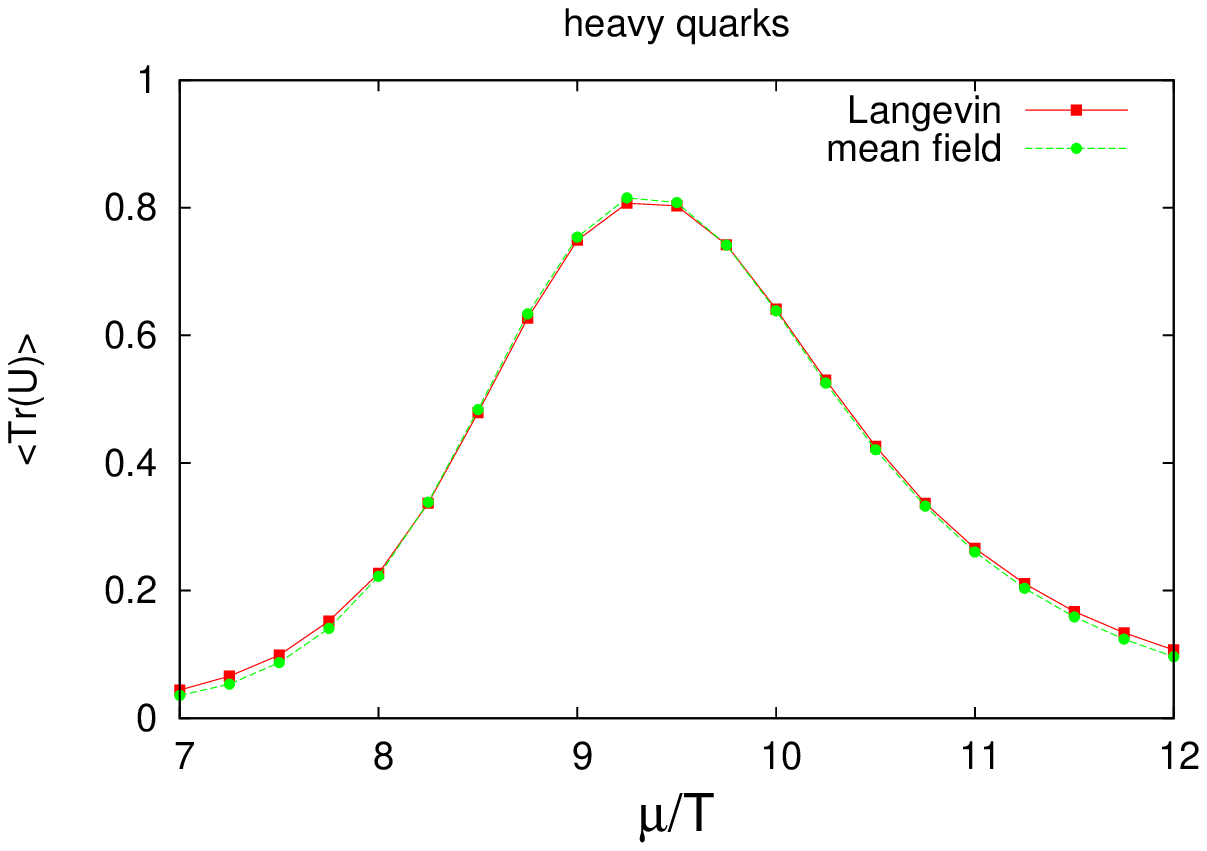}
}
\subfigure[~$\langle \tr(U^\dg)\rangle$]  % caption for subfigure a
{   
 \label{vhq}
 \includegraphics[scale=0.6]{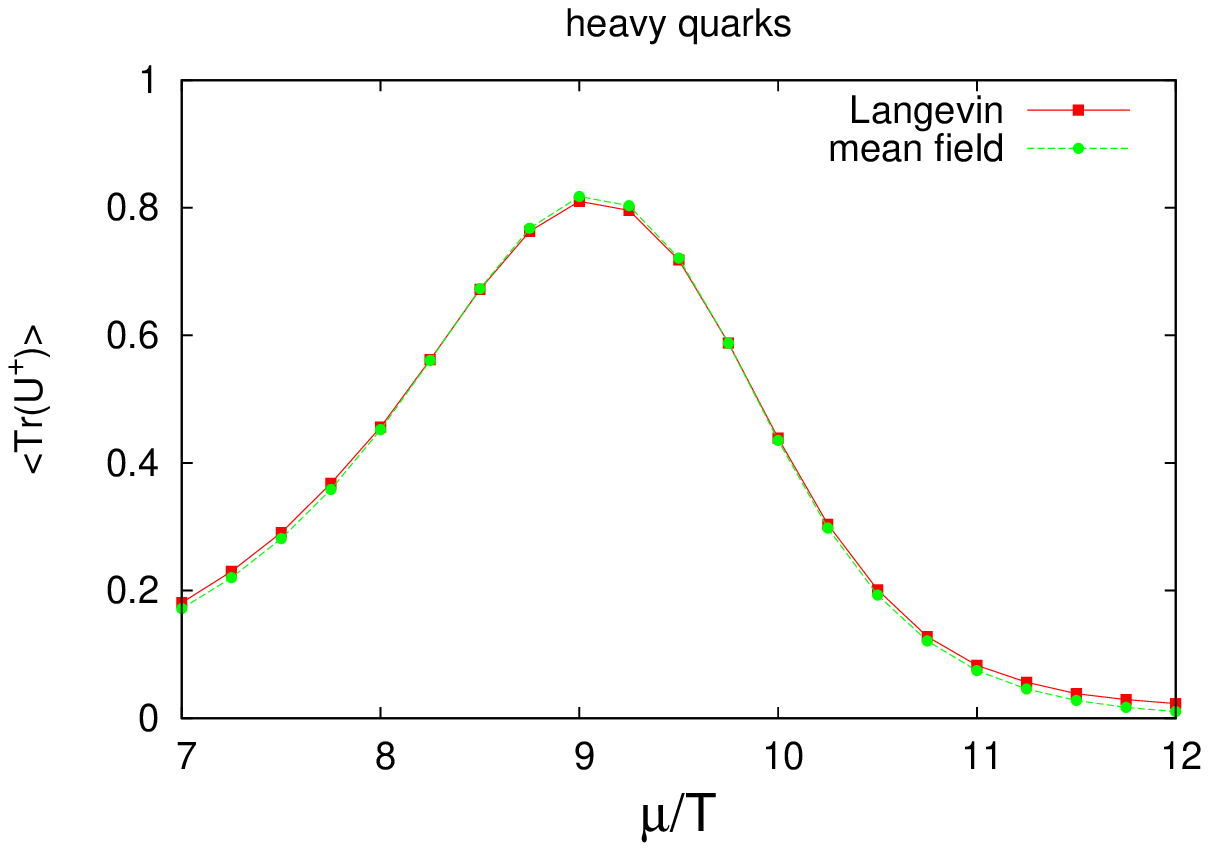}
}
\subfigure[~density]  % caption for subfigure a
{   
 \label{nhq}
 \includegraphics[scale=0.6]{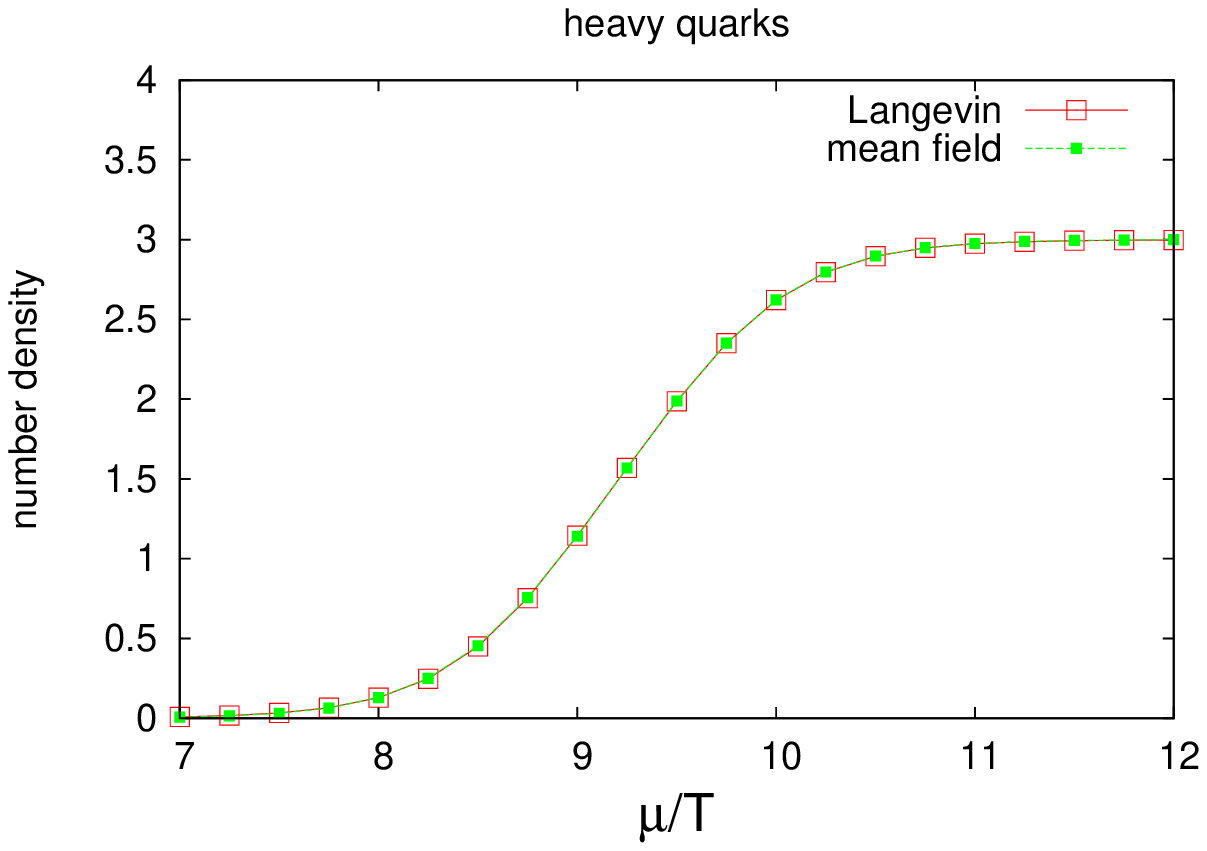}
}
\caption{Comparison of Polyakov lines $\langle \tr(U) \rangle, \langle \tr(U^\dg) \rangle$ and number density vs.\ $\mu/T$, computed via  complex Langevin and mean field techniques in the heavy quark model. Note the saturation at high $\m/T$
at density=3.}
\label{hq}
\end{figure}

    The second example is the heavy quark model described in the previous section.  Figure  \ref{hq}
shows the comparison plot of $\langle \tr[U_\vx] \rangle, \langle \tr[U^\dg_\vx] \rangle$ and number density obtained from the complex Langevin and mean field techniques.  Once again, the data points obtained from each technique are hardly distinguishable.   The saturation of number density at density=3 is the value expected from the Pauli principle for staggered fermions.

\begin{figure}[htb]
\centerline{\scalebox{0.6}{\includegraphics{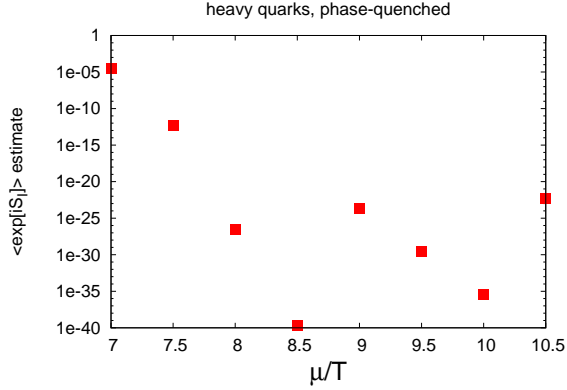}}}
\caption{An estimate of $\langle \exp[iS_I] \rangle_{pq}$ vs.\ $\m/T$ in the phase-quenched version of the heavy quark model, obtained from the second order cumulant.  $S_I$ is the imaginary part of the action.}
\label{phasehq}
\end{figure}

   The second-cumulant estimate for $\langle e^{iS_I} \rangle_{pq}$ vs.\ $\m/T$ is shown in Fig.\ \ref{phasehq}.  The sign problem is severe, although not as severe as in the gauge-Higgs example at higher chemical potentials.  

\begin{figure}[htb]
\includegraphics[scale=0.6]{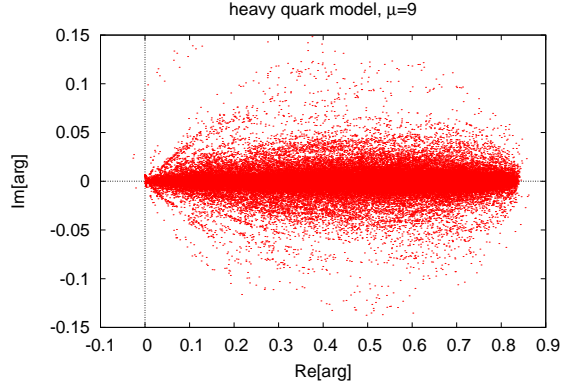}
\caption{Argument of the logarithm for the heavy quark model at chemical potential
$\m/T=9$, evaluated at each Langevin time step.  As in the previous example, values near the negative real axis are a negligible fraction of the sample.}
\label{dethq}
\end{figure}

   When the complex Langevin equation is applied to the heavy quark model, the action contains the logarithm of a product of integration measure and determinant factors at each site $\vx$ on the lattice
\bea
\mbox{Arg} &=& \sin^2\left({\th_1(\vx)-\th_2(\vx) \over 2}\right)\sin^2\left({2\th_1(\vx)+\th_2(\vx) \over 2}\right)\sin^2\left({\th_1(\vx)+2\th_2(\vx) \over 2}\right)
\non \\
& & \times \big(1 + he^{-\m/T} \tr[U_\vx^\dg] + h^2 e^{-2\m/T} \tr[U_\vx] + h^3 e^{-3\m/T} \big)
\non \\
& & \times \big(1 + he^{\m/T} \tr[U_\vx] + h^2 e^{2\m/T} \tr[U_\vx^\dg] + h^3 e^{3\m/T} \big) \ .
\eea
Once again we must monitor the argument of the logarithm, to check that Langevin evolution does not entail frequent crossings of the branch cut on the negative real axis.  A plot of values for Arg obtained at each Langevin time step, at the chemical potentials $\m/T=9$, is shown in Fig.\ \ref{dethq}.  As in the previous gauge-Higgs example, the argument of the logarithm is a product of factors at each lattice site, and Fig.\ \ref{hq}  displays the value, in the complex plane, of a particular factor associated with an arbitrarily selected lattice site $\vx=\vx'$.  The values seem to be strongly concentrated in a region with where the real part of the value is positive, so crossings of the branch cut do not appear to be of concern in this case either.

\subsection{Gauge-Higgs at $\k=3.9$}

\begin{figure}[htb]
\subfigure[~$\langle \tr(U)\rangle$]  % caption for subfigure a
{   
 \label{uhq}
 \includegraphics[scale=0.6]{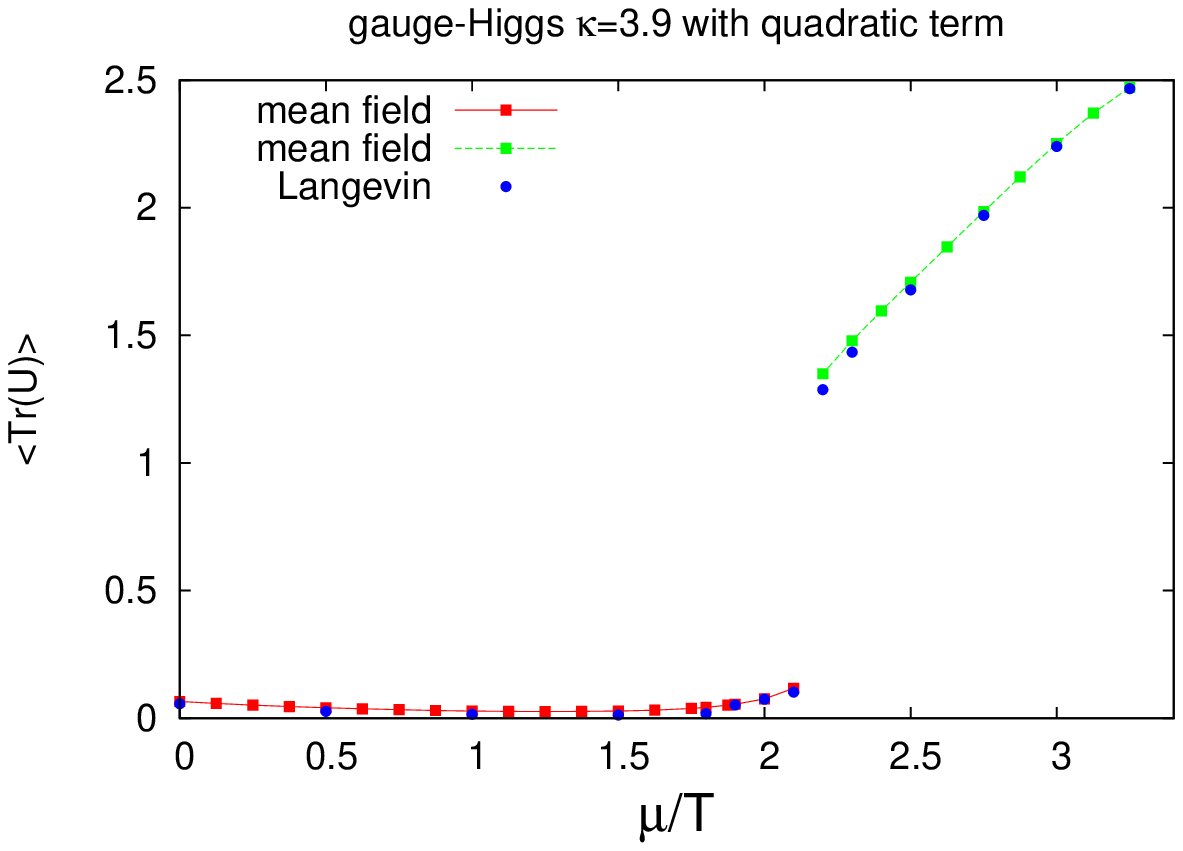}
}
\subfigure[~$\langle \tr(U^\dg)\rangle$]  % caption for subfigure a
{   
 \label{vhq}
 \includegraphics[scale=0.6]{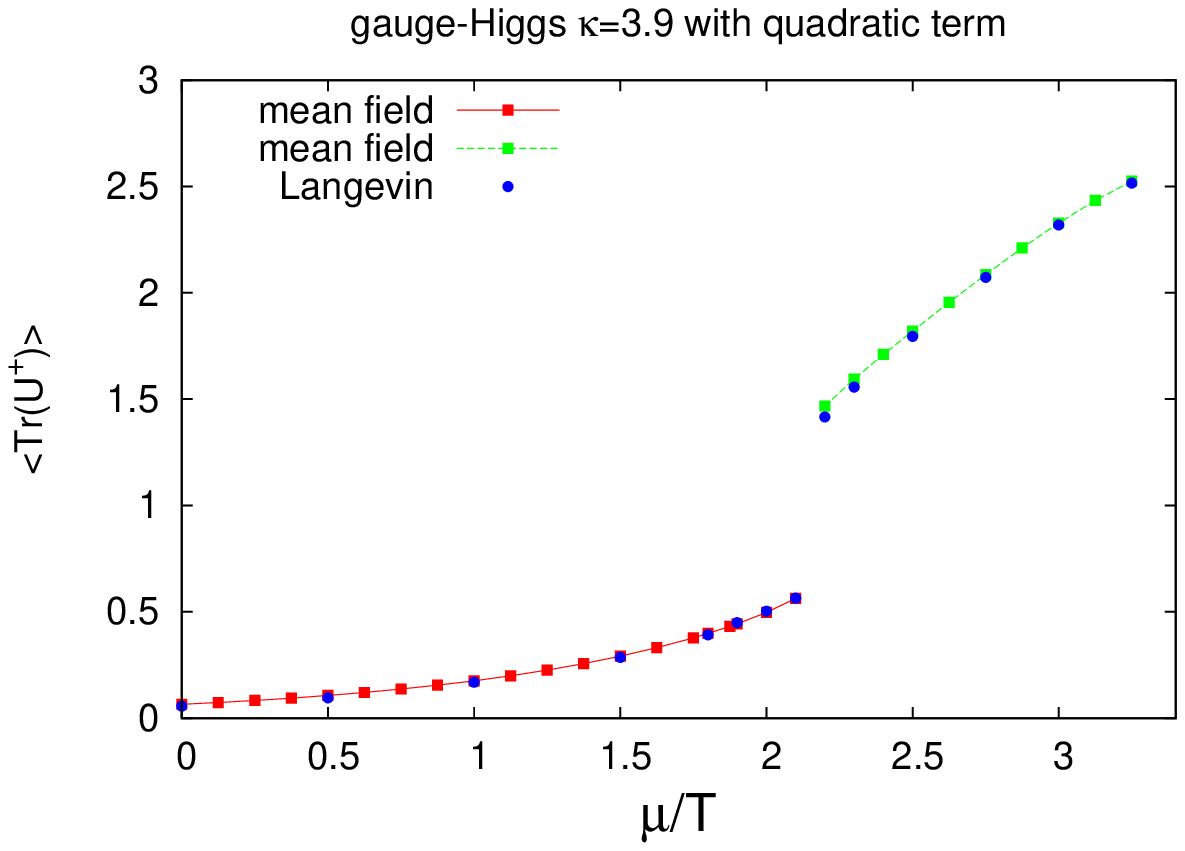}
}
\subfigure[~density]  % caption for subfigure a
{   
 \label{nhq}
 \includegraphics[scale=0.6]{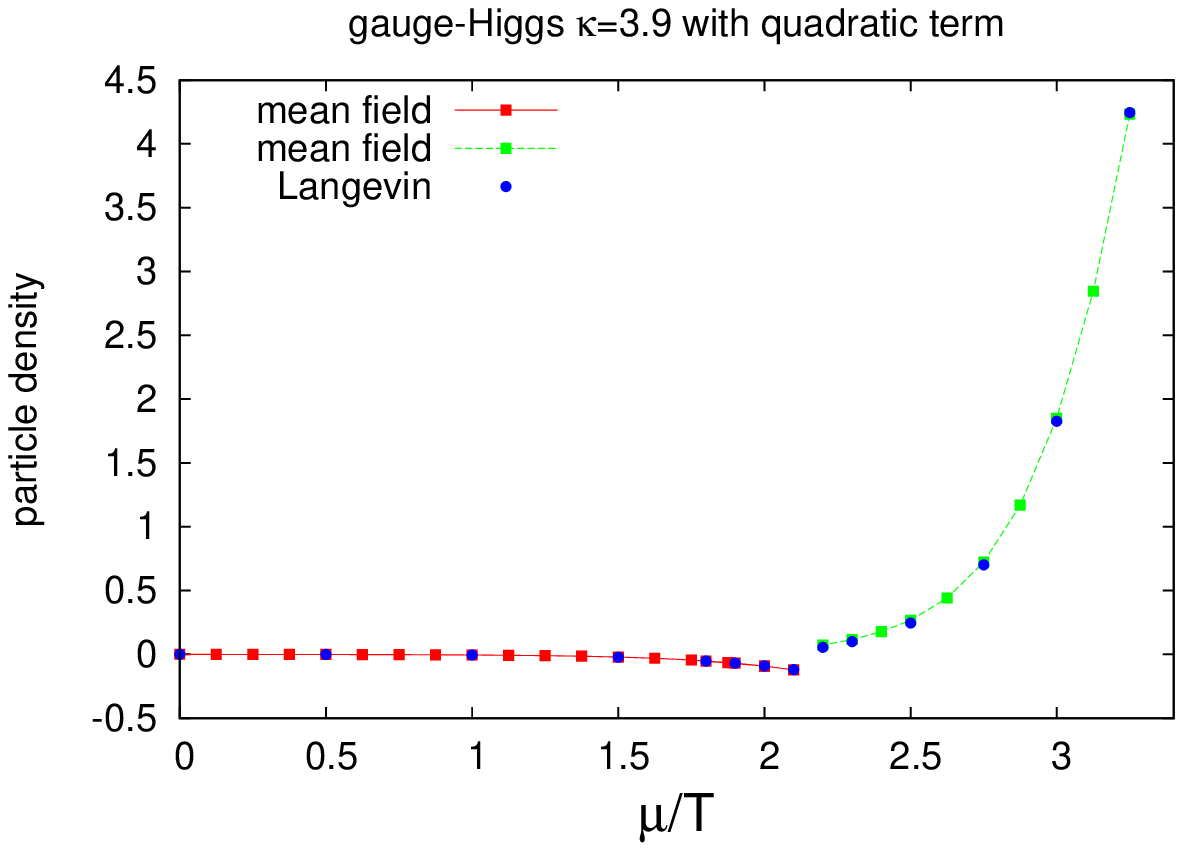}
}
\caption{Comparison of Polyakov lines $\langle \tr(U) \rangle, \langle \tr(U^\dg) \rangle$ and number density vs.\ $\m/T$, computed via  complex Langevin and mean field techniques, in gauge-Higgs theory at $\kappa=3.9$ for the action $S_P$ in eq.\
\rf{SP39q}, which includes quadratic center symmetry-breaking terms.}
\label{gh39q}
\end{figure}

   So far we have not seen any evidence of a phase transition.  We now consider the gauge-Higgs theory at $\b=5.6, \k=3.9$
on a $16^3 \times 6$ lattice volume. At $\k=3.9$ the gauge-Higgs system is closer to the confinement-like to Higgs-like crossover, at around $\k=4.0$, and therefore this corresponds to a lighter scalar particle, as compared to $\k=3.8$.  The effective action, including the local quadratic term, was given in \rf{SP39q}.  The mean field and complex Langevin results for $\langle \tr[U_\vx] \rangle, \langle \tr[U^\dg_\vx] \rangle$ and number density are shown
in Fig.\ \ref{gh39q}. The apparent discontinuity in all three observables strongly suggests a first-order phase transition at a value of $\m$ between 2.1 and 2.2, so we see that, in comparison to the previous two examples, a transition emerges at lighter particle masses.  Once again, the difference between the mean field and complex Langevin results is barely discernable, and both methods agree on the position of the transition.  In the case of mean field there are two solutions of eq.\ \rf{cle} in the neighborhood of the transition, with free energies almost identical at the transition.  Above and below the transition one chooses the solution with the lowest free energy

   One point that is worth noting is that at the higher $\m$ values there is also more than one solution of the complex Langevin equation, and which solution is chosen by the system depends on the starting point of the evolution.  We consider two initializations at $t=0$ in Langevin time:

\begin{description}
{     \item[\bf I]~~ $\th_1(x)=\th_2(x)=0 , ~ \tr[U_x]=3$.
     \item[\bf II]~~  $\th_1(x)=-\th_2(x)={2\pi \over 3} , ~ \tr[U_x]=0$.  }
\end{description}

\begin{figure}[htb]
\subfigure[~$\m/T=1.9$]  % caption for subfigure a
{   
 \label{split1p9}
 \includegraphics[scale=0.6]{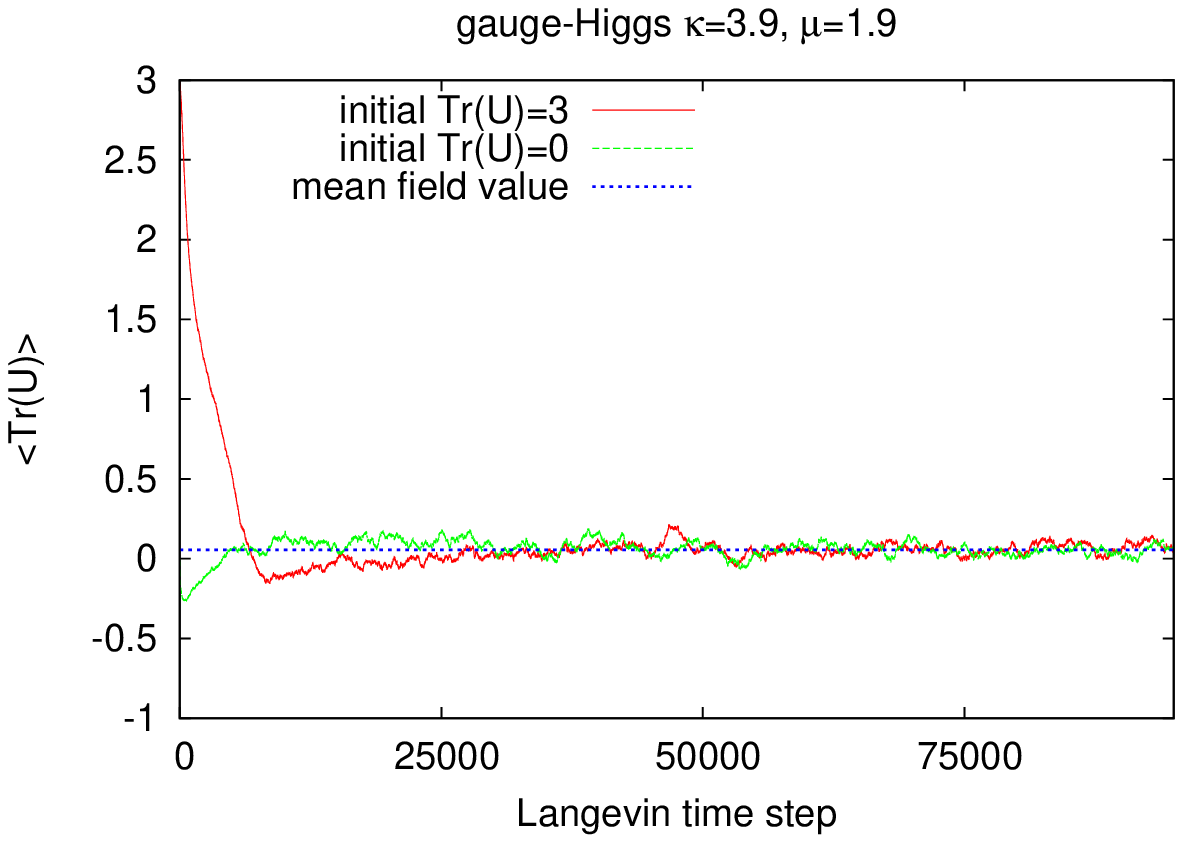}
}
\subfigure[~$\m/T=3.0$]  % caption for subfigure a
{   
 \label{split3p0}
 \includegraphics[scale=0.6]{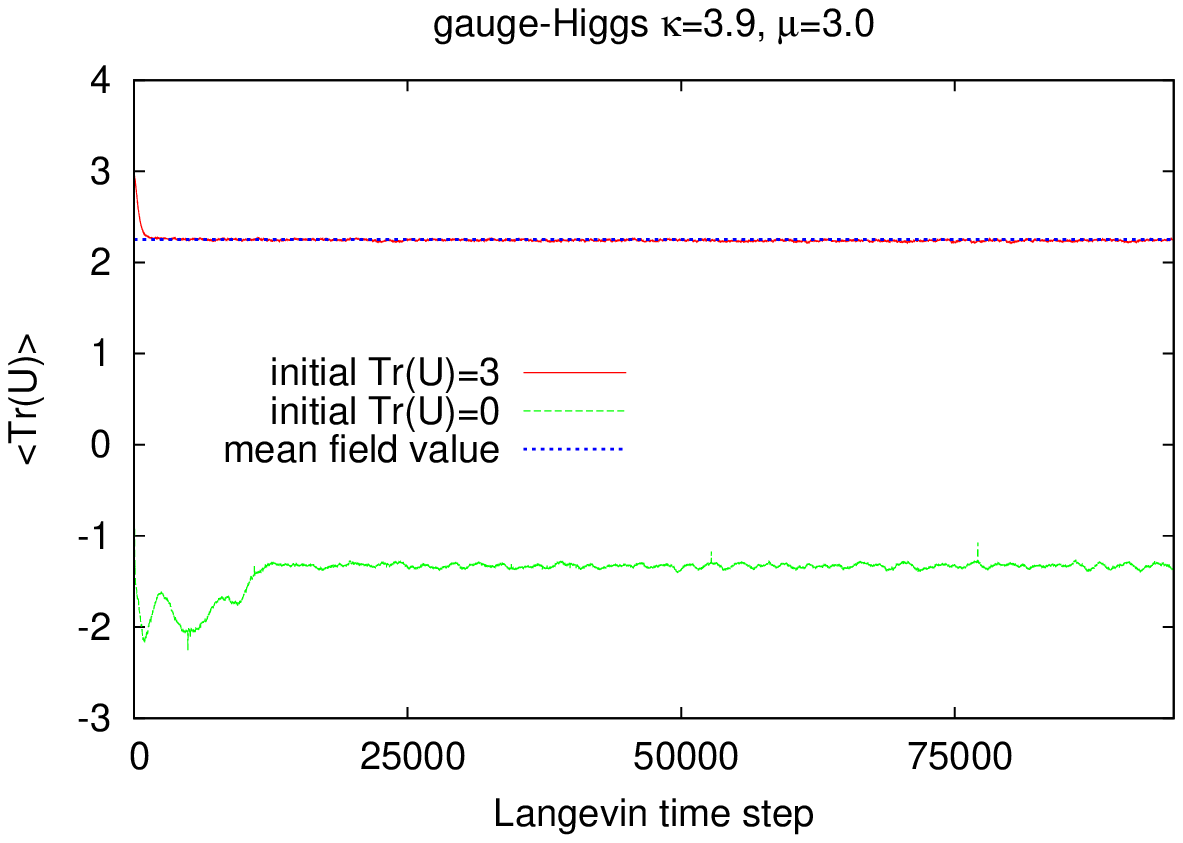}
}
\caption{Dependence of Langevin evolution on initial conditions, in the gauge-Higgs model with a quadratic symmetry-breaking term. (a) Initial conditions lead to convergent Langevin evolution, in agreement with the mean field solution, at $\m/T=1.9$.
(b) Initial conditions lead to two different solutions of the Langevin equation at $\m/T=3.0$.  The upper solution is in close agreement with mean field theory.}
\label{split}
\end{figure}

At low values of the chemical potential, the choice of initialization doesn't matter; both solutions converge to the same values,
as seen in a plot (Fig.\ \ref{split1p9}) of the lattice volume average of $\tr(U_x)$ at each Langevin time step (no average over time), at $\m=1.9$
At higher values of $\m$, the two initializations lead to different solutions, seen in Fig.\  \ref{split3p0} at $\m=3.0$.  The upper solution, with initialization I, agrees very well with mean field theory, and in fact this initialization is used for the Langevin data
shown in Fig.\ \ref{gh39q}.  Then there is a question of why should we prefer the solution I, which agrees with mean field, rather than solution II, which disagrees with mean field. 

\begin{figure}[htb]
\subfigure[~initialization I]  % caption for subfigure a
{   
 \label{detAq}
 \includegraphics[scale=0.6]{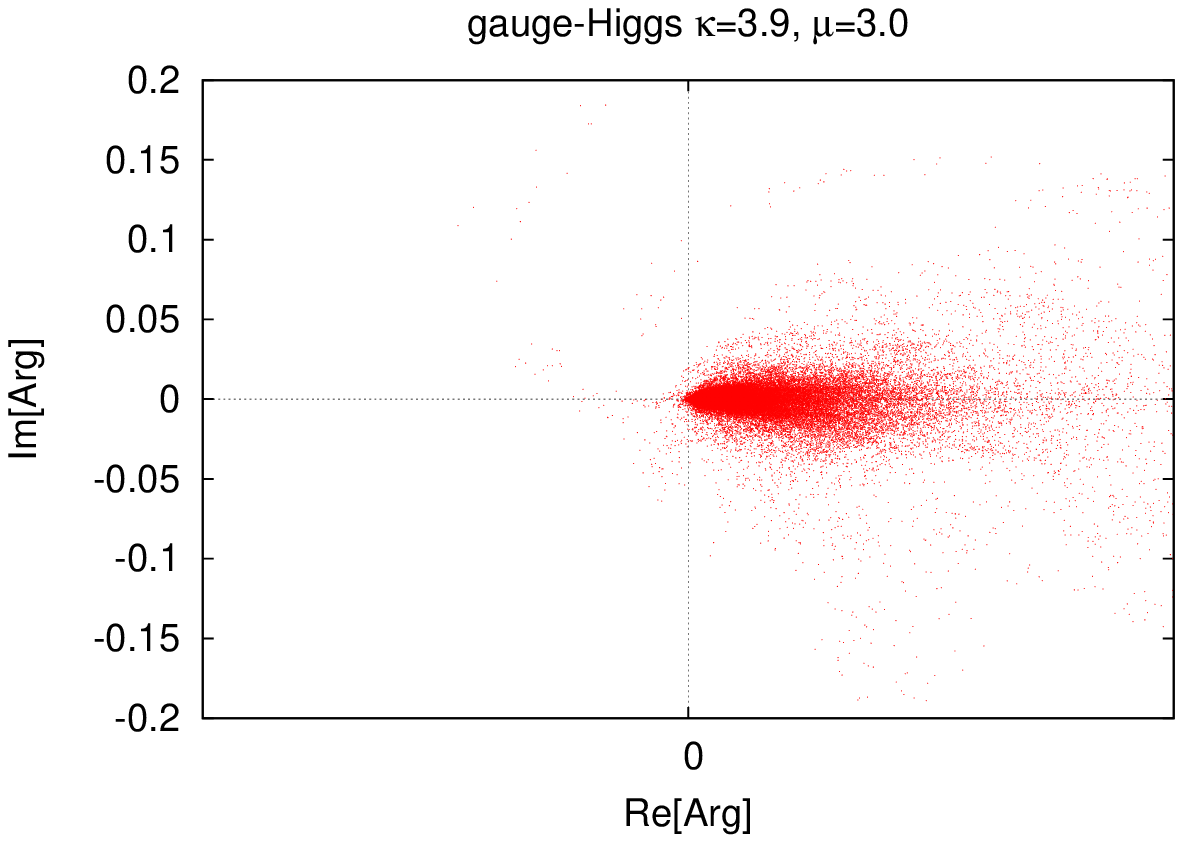}
}
\subfigure[~initialization II]  % caption for subfigure a
{   
 \label{detBq}
 \includegraphics[scale=0.6]{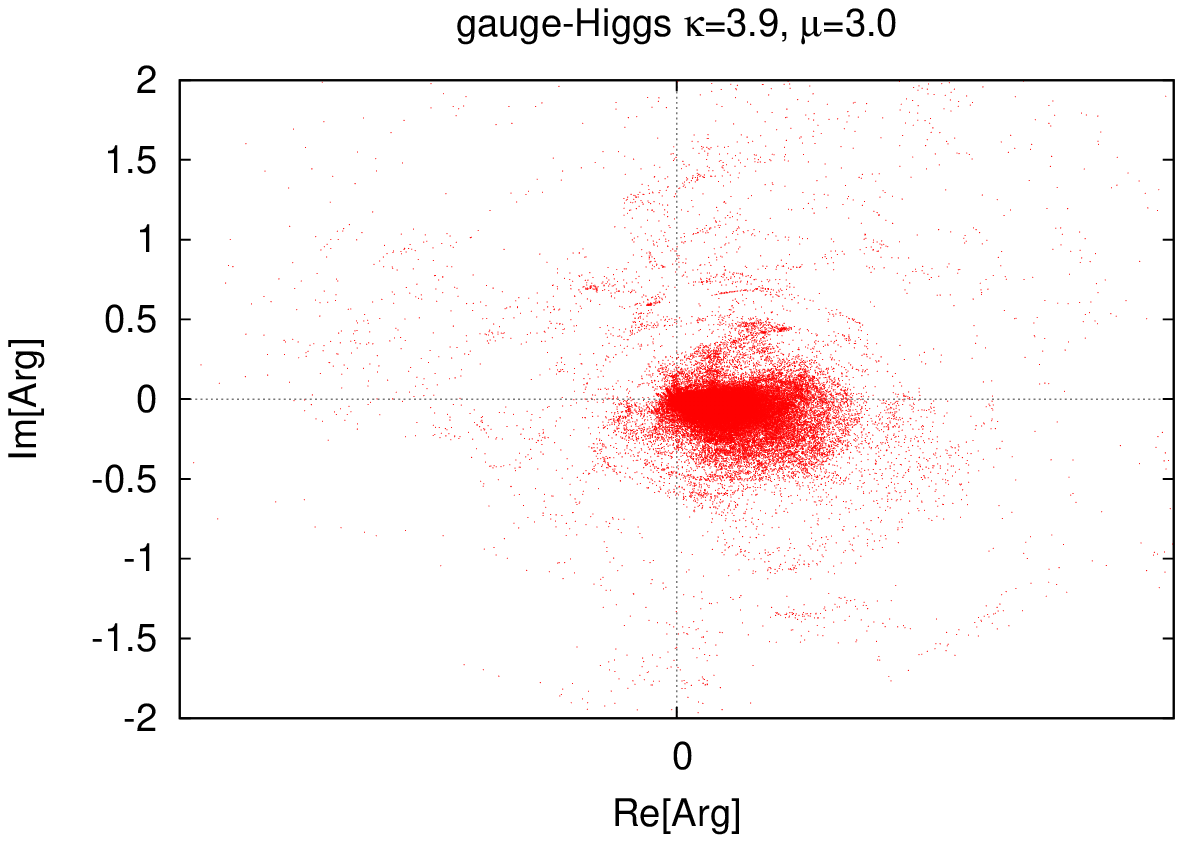}
}
\caption{Argument of the logarithm at $\m/T=3.0$ in the gauge-Higgs model with a center symmetry-breaking quadratic term,
for two different initializations of the Langevin evolution.  (a) Initialize with $\tr U=3$.  This leads to the upper solution of Langevin evolution in Fig.\ \ref{split3p0}.  (b) Initialize with $\tr U=0$.  This leads to the lower solution of Langevin evolution in Fig.\ \ref{split3p0}.}
\label{detABq}
\end{figure}

\begin{figure}[htb]
\subfigure[~solution I]  % caption for subfigure a
{   
 \label{theta1a}
 \includegraphics[scale=0.6]{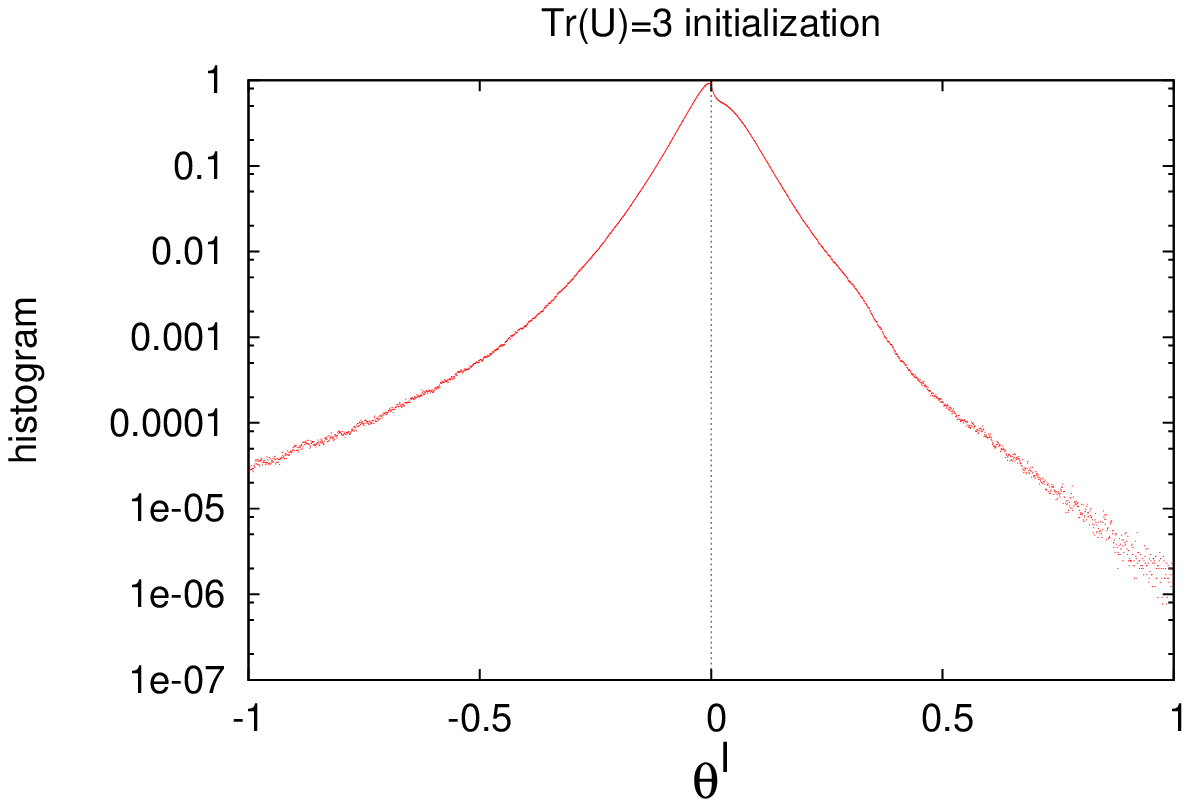}
}
\subfigure[~solution II]  % caption for subfigure a
{   
 \label{theta1b}
 \includegraphics[scale=0.6]{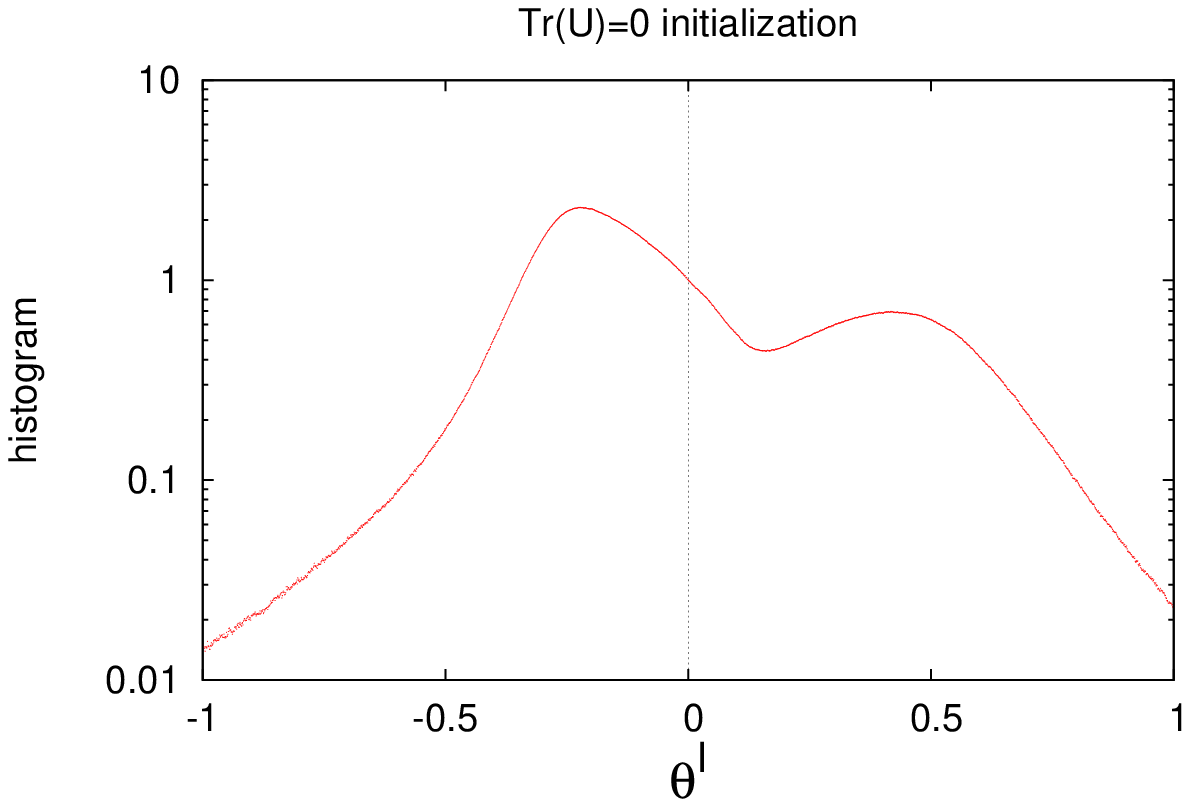}
}
\caption{Histogram of the imaginary part $\th^I$ of the degrees of freedom $\th_{1,2}(\vx)$, normalized to unity at $\th^I=0$.
(a) Initialize with $\tr U=3$, leading to solution I in agreement with mean field; (b) initialize with $\tr U=0$, leading to solution II in disagreement with mean field.  Note the logarithmic scales in each plot, and that for solution II the angle distribution is broad and not very well localized in the complex plane.}
\label{theta}
\end{figure}

    There are two reasons.  First, at the higher values of $\m$ where solution II differs from mean field,
solution II is invalidated by a branch cut problem.  Fig.\ \ref{detABq} is a plot of the argument of the logarithm at an arbitrary site on the lattice, at each Langevin time step, for initial conditions I and II at $\m/T=3.0$.  For the solution which develops from initial conditions I, the argument of the logarithm is mostly well away from the branch cut on the negative real axis.  This is much less so for solution II, where there are many more points near the negative real axis and, as a consequence, there must be many
crossings of the branch cut in Langevin evolution.  This suggests a branch cut crossing problem in solution II, so solution I is preferred.  The second reason for preferring solution I concerns the probability distribution of the degrees of freedom in the complex plane.  It is well known that the complex Langevin approach can fail if this probability distribution is not sufficiently well localized in the complex plane \cite{Aarts:2013uza}. Following ref.\ \cite{Aarts:2011zn}, we make a histogram of the distribution
obtained for the imaginary part of the $\th_{1,2}$ angles at $\m/T=3$, with the initialization $\tr U=3$, leading to solution I, and
initialization $\tr U=0$, leading to solution II.  The two histograms, arbitrarily normalized to unity at $\th_I=0$, are shown in Fig.\
\ref{theta}.  We see that the distribution of the imaginary part $\theta_I$ of the $\th_{1,2}$ angles is well localized for solution I,
and very broad and not well-localized for solution II, which indicates that the latter solution produces incorrect results.

\clearpage

\subsection{Gauge-Higgs at $\k=3.9$, quadratic symmetry-breaking terms neglected}

   For the last example we consider the action \rf{SP39}, which follows from \rf{SP39q} using the identities \rf{identities} and dropping terms proportional to $\tr[U_\vx]^2$ and $\tr[U^\dg_\vx]^2$.

\begin{figure}[htb]
\subfigure[~$\langle \tr(U) \rangle$]  % caption for subfigure a
{   
 \label{u39}
 \includegraphics[scale=0.6]{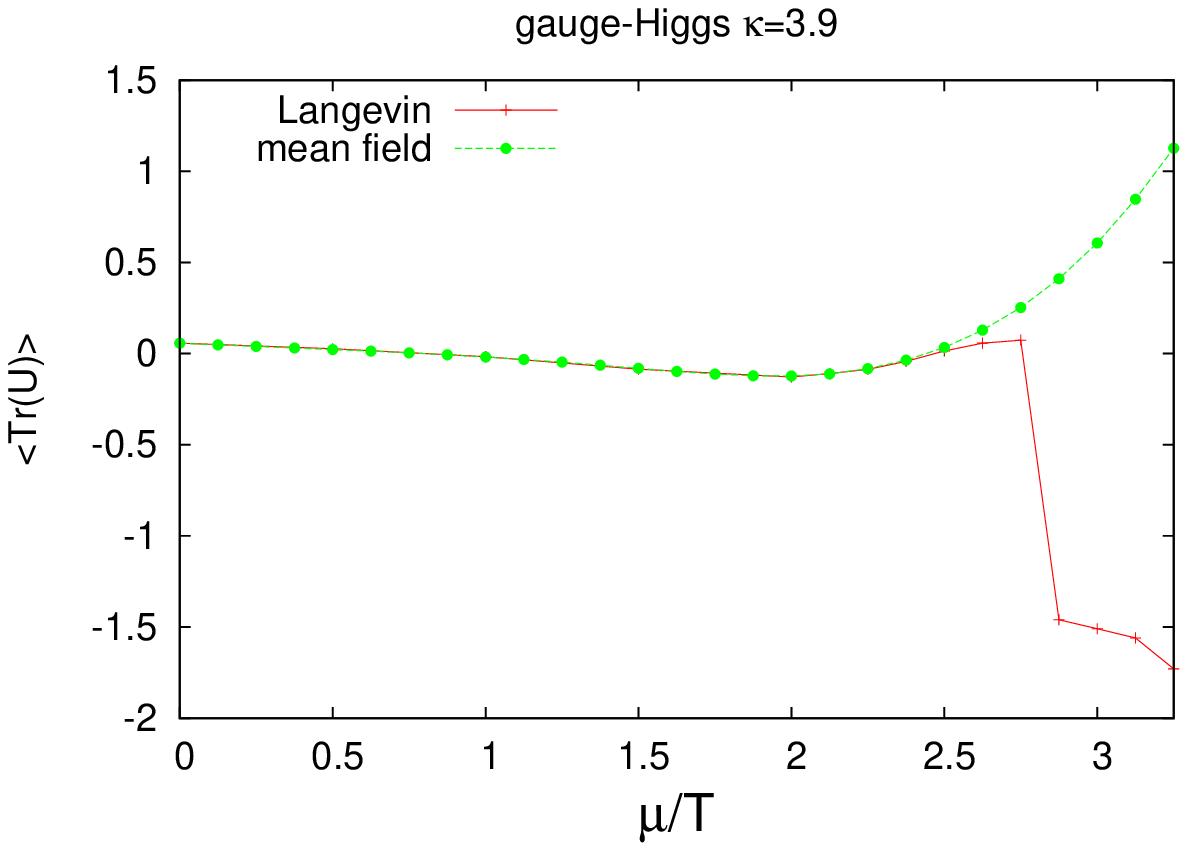}
}
\subfigure[~$\langle \tr(U^\dg) \rangle$]  % caption for subfigure a
{   
 \label{v39}
 \includegraphics[scale=0.6]{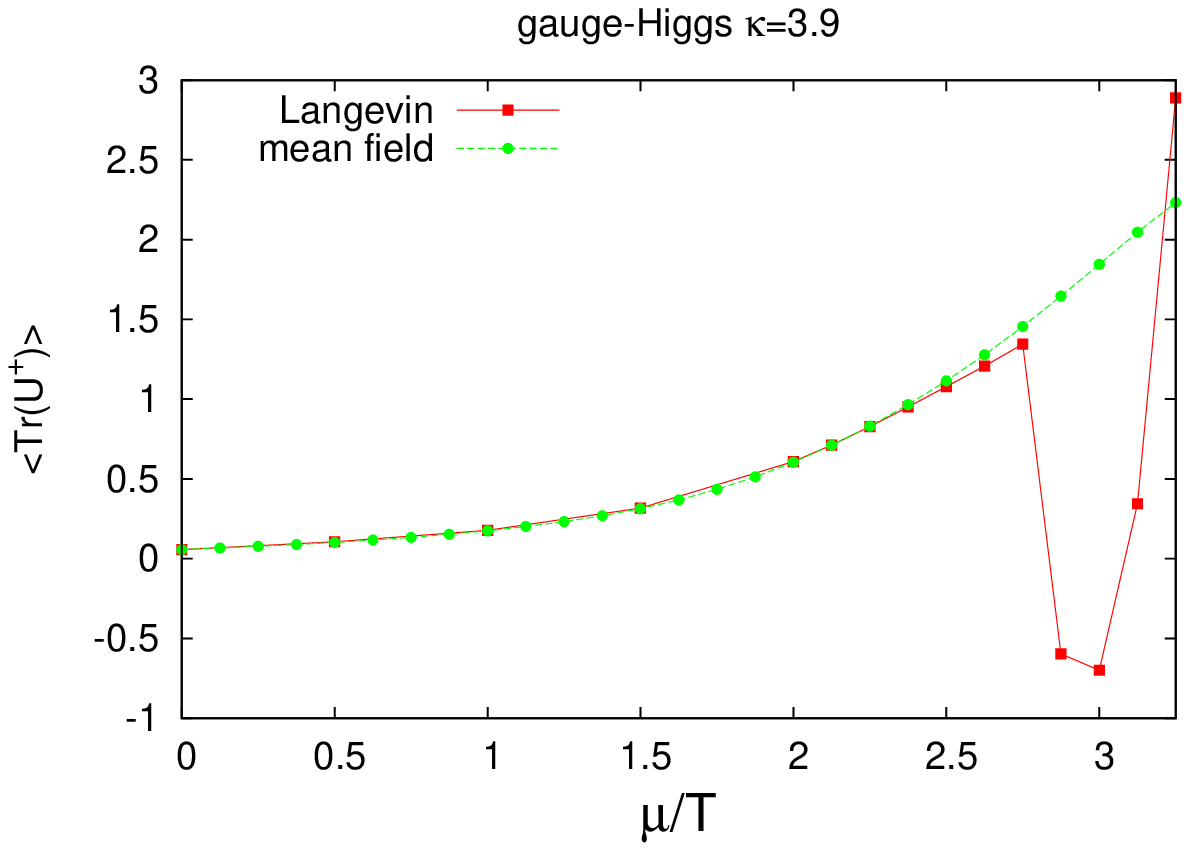}
}
\subfigure[~density]  % caption for subfigure a
{   
 \label{n39}
 \includegraphics[scale=0.6]{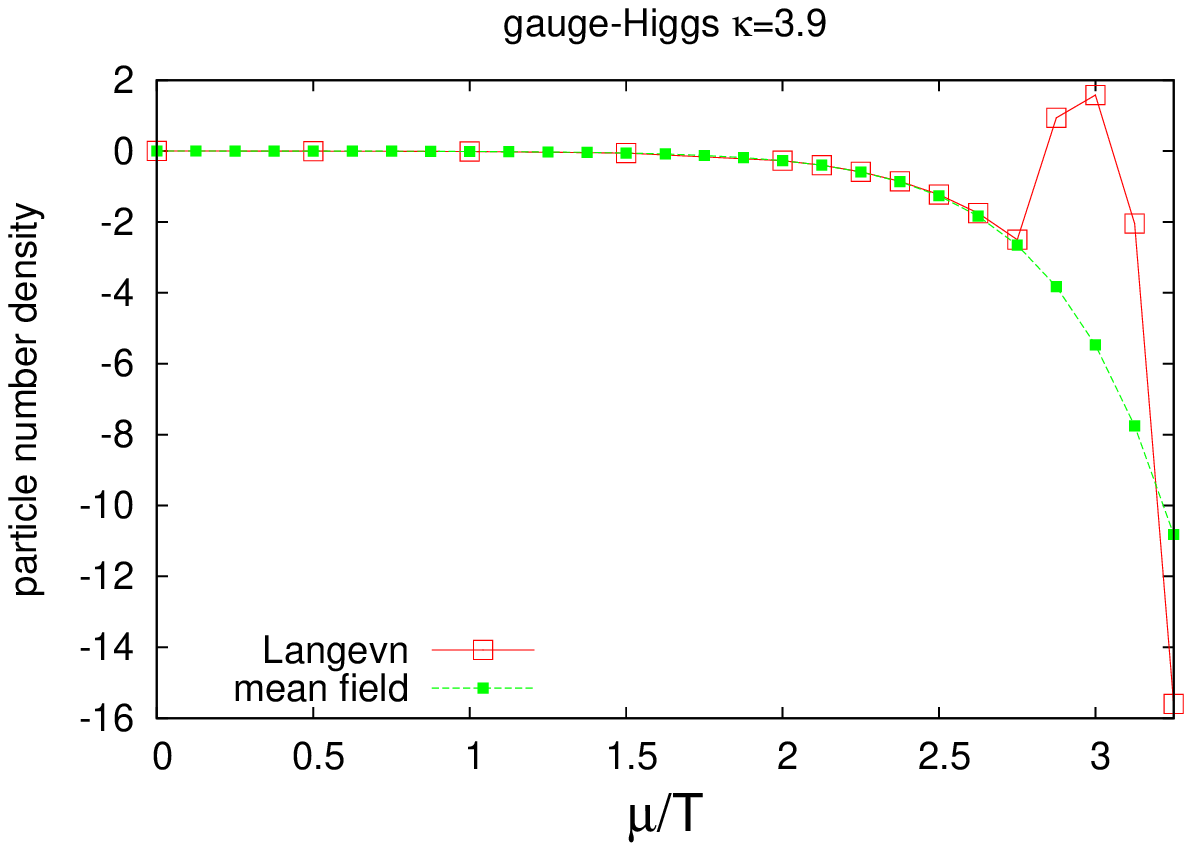}
}
\caption{Comparison of Polyakov lines $\langle \tr(U) \rangle, \langle \tr(U^\dg) \rangle$ and number density vs.\ $\m/T$, computed via  complex Langevin and mean field techniques, in gauge-Higgs theory at $\kappa=3.9$ for the action $S_P$ in eq.\
\rf{SP39}, where quadratic center symmetry-breaking terms are neglected.}
\label{gh39}
\end{figure}

\begin{figure}[htb]
\subfigure[~$\m/T=1.5$]  % caption for subfigure a
{   
 \label{}
 \includegraphics[scale=0.6]{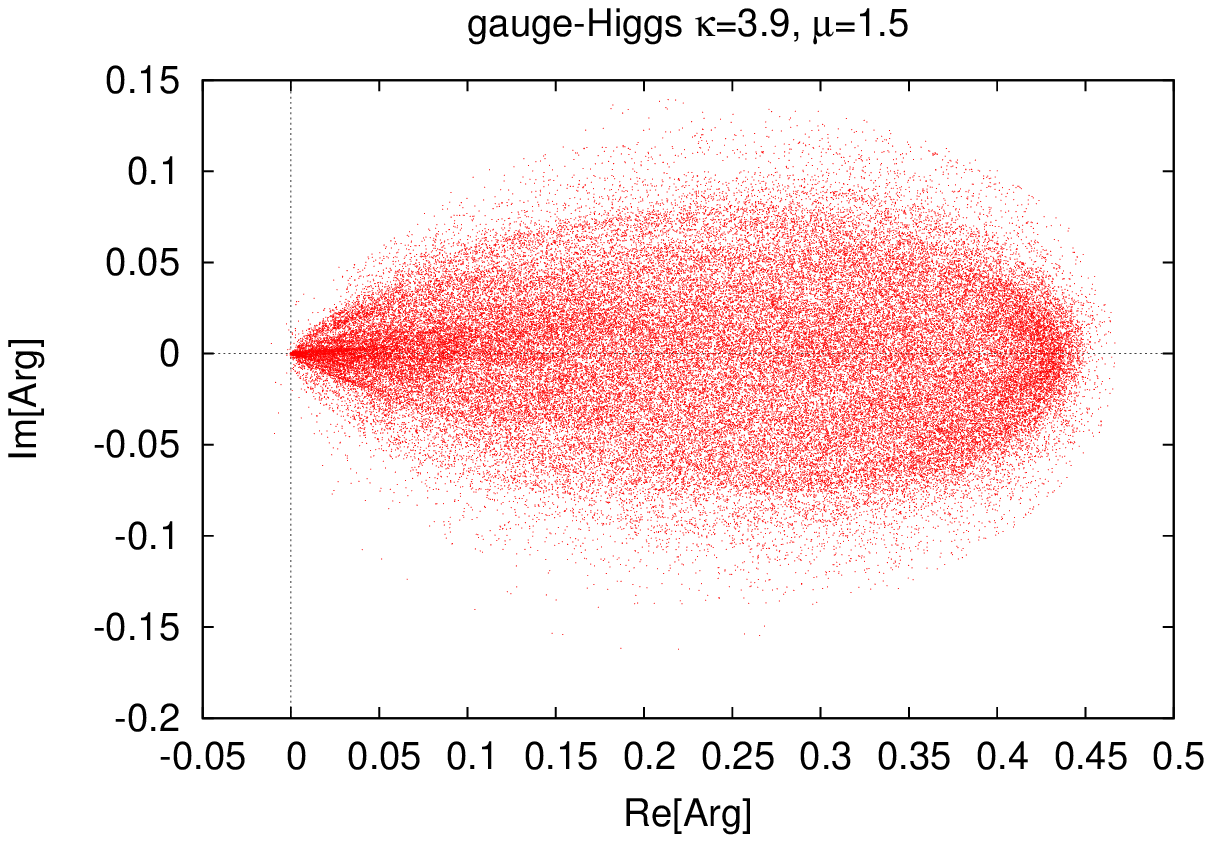}
}
\subfigure[~$\m/T=2.0$]  % caption for subfigure a
{   
 \label{}
 \includegraphics[scale=0.6]{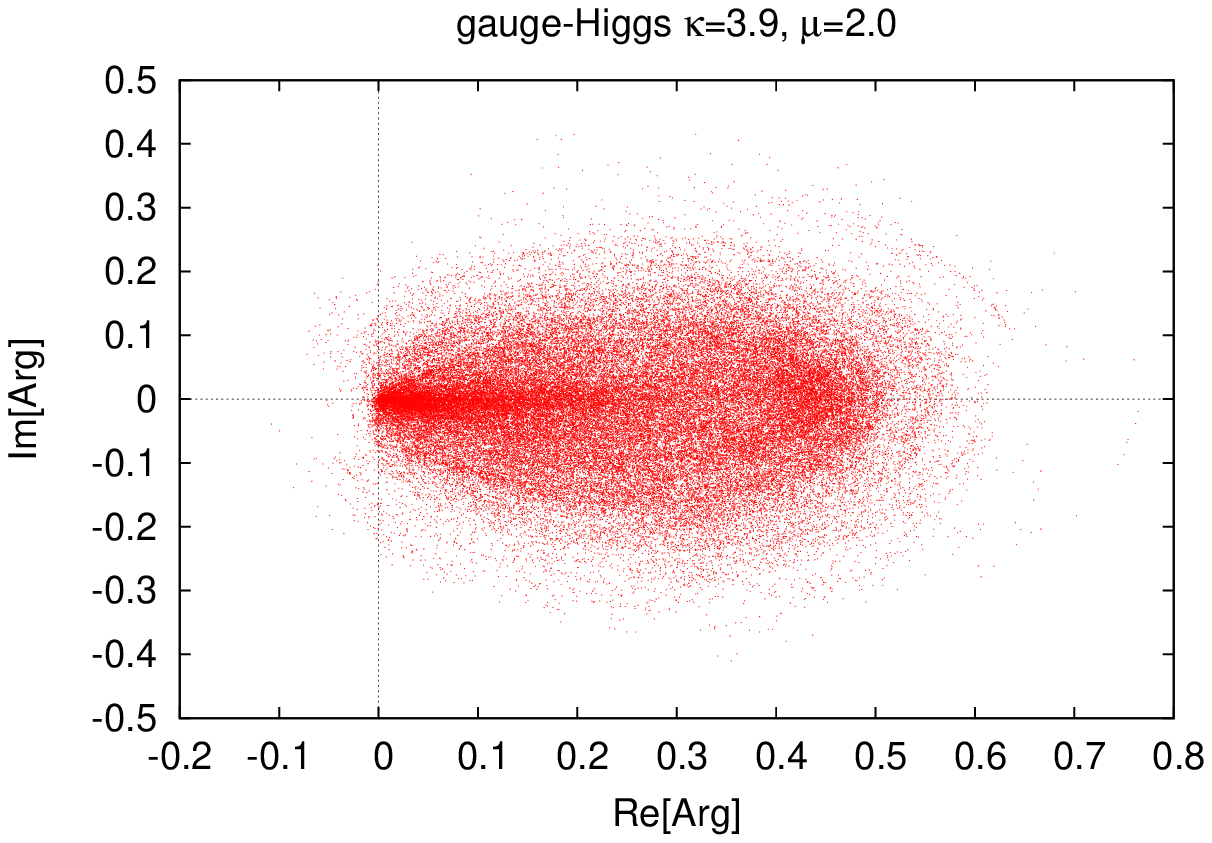}
}
\subfigure[~$\m/T=2.75$]  % caption for subfigure a
{   
 \label{}
 \includegraphics[scale=0.6]{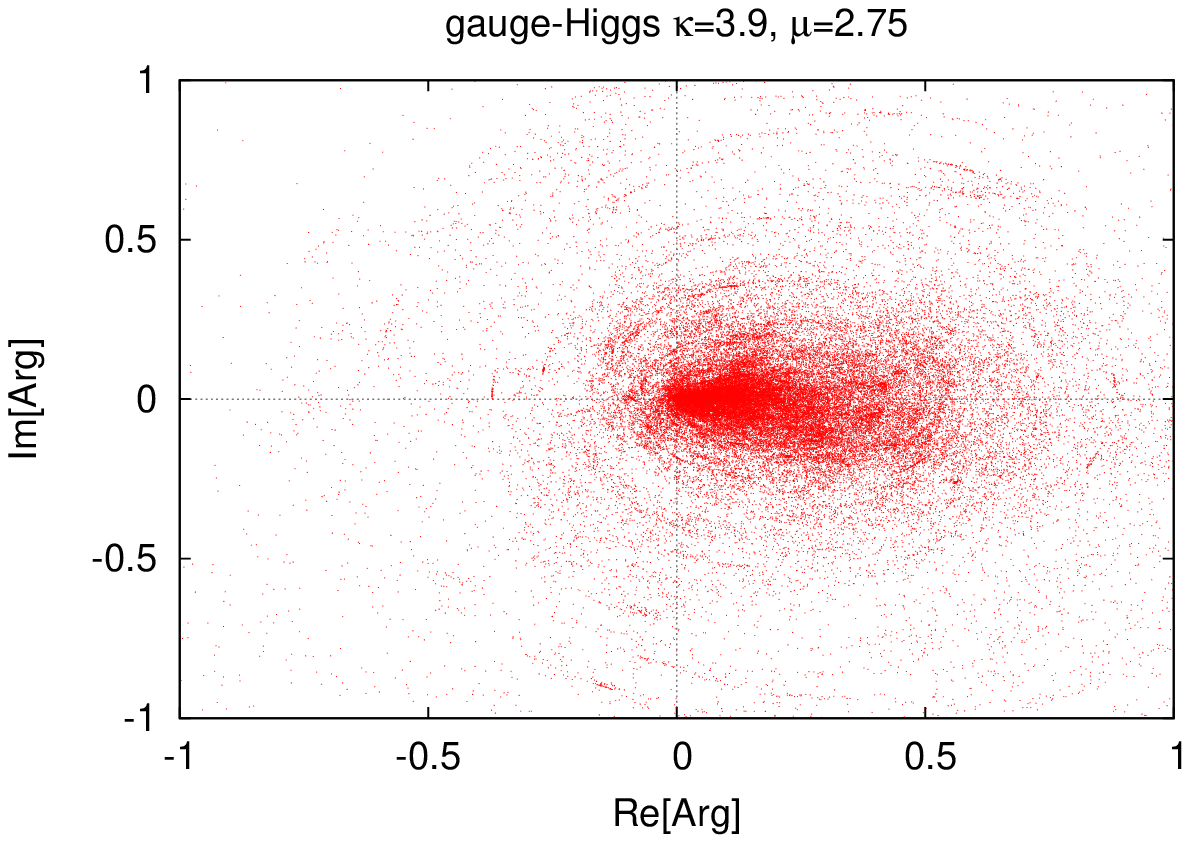}
}
\subfigure[~$\m/T=3.25$]  % caption for subfigure a
{   
 \label{}
 \includegraphics[scale=0.6]{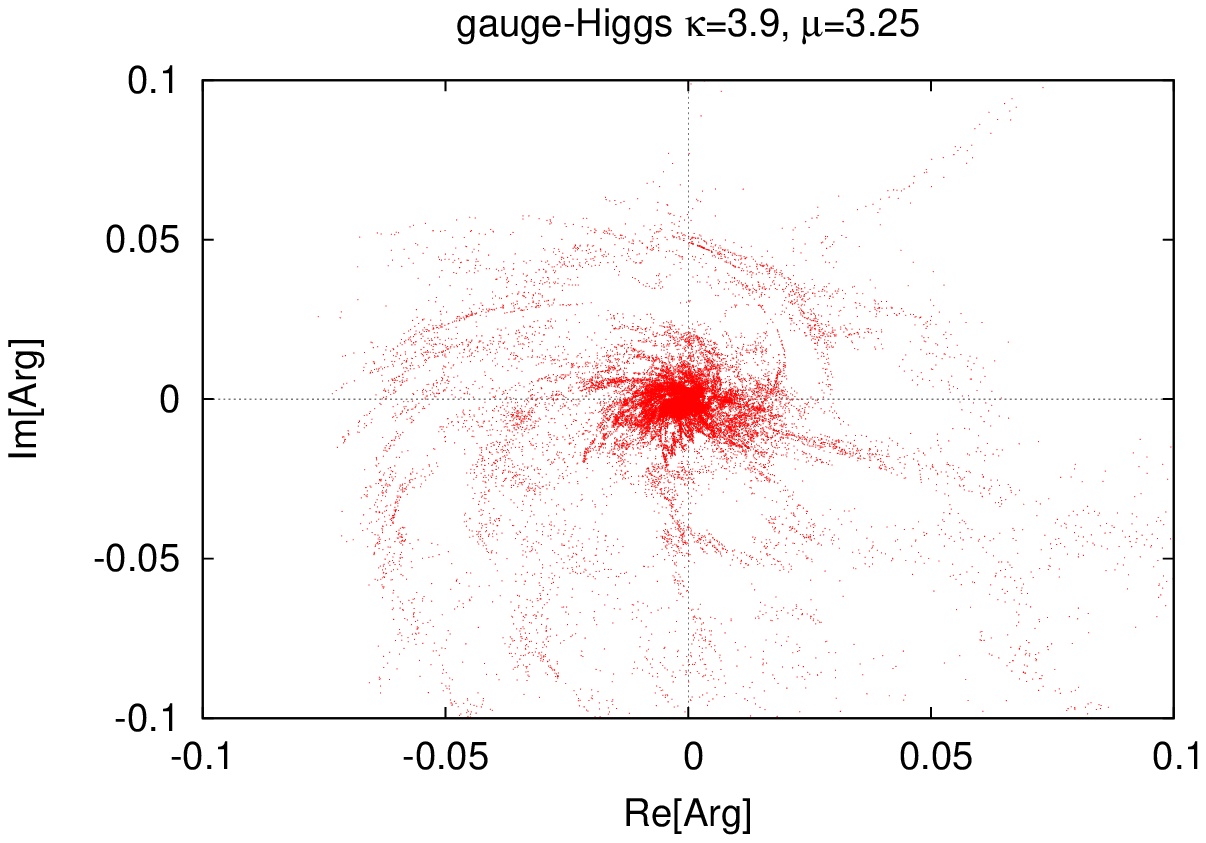}
}
\caption{Argument of the logarithm for gauge-Higgs theory at $\b=5.6,~\k=3.9$, and chemical potentials
$1.5 \le \m/T \le 3.25$ (subfigures a-d), evaluated at each Langevin time step.  The presence of many points near the negative real axis is very plain at $\b \ge 2.75$, signaling the presence of a branch cut problem.}
\label{detgh39}
\end{figure}

   In contrast to the previous three examples, while the complex Langevin and mean field methods agree quite closely for
$\langle \tr U \rangle, \langle \tr U^\dg \rangle$ and particle density up to 
$\mu \approx 2.75$, they give very different answers at $\m>2.75$.  The explanation of this discrepancy is explained
by a plot (Fig.\ \ref{detgh39}) of the argument of the logarithm in the action, eq.\ \rf{measure}.  At the lower values of $\m$, the values of the argument are well away from the negative real axis, and we deduce that crossing the logarithmic branch cut is a rare event.
However, at $\mu \ge 2.75$, there are many values of the argument which lie close to the negative real axis, implying that
Langevin evolution will commonly cross the logarithmic branch cut.  In that case, we can no longer trust the results derived from the complex Langevin equation.   Then the question is whether, by altering the initial conditions, one may obtain a solution of
the Langevin equation which does not have a branch cut problem.  It is impossible to explore all initial conditions, of course.
Initial conditions I and II have been tried, with $\tr[U_x]=3$ and $\tr[U_x]=0$, respectively, along with an intermediate initialization 
\begin{description}
\item[\bf III] ~~$\theta_1(\vx) = 0.3 \pi r_1(\vx) ~~~,~~~  \theta_2(\vx) = 0.3 \pi r_2(\vx)$
\end{description}
where $r_1(\vx),r_2(\vx)$ are linearly distributed random numbers in the range $[0,1]$.  All three initializations run into a branch cut problem around $\m=2.75$.

    One other result in this example, seen in both the mean field and Langevin solutions, is that the particle number density  displayed in Fig.\ \ref{n39} can become large and negative at large $\m$, a result which we consider unphysical since the chemical potential is positive.  This is clear evidence that, although the neglected quadratic symmetry-breaking terms may have a relatively small effect on correlators at $\m=0$, they cannot be ignored at finite $\m$.  We have neglected them in this example only for the purpose of comparing Langevin and mean field results for another action belonging to the class of SU(3) spin models.
   
\section{Conclusions}

   There are two main results.  The first is that  where the complex Langevin and mean field results agree, in the cases studied so far, they agree to an extraordinary degree of accuracy.  It is natural to ask why these mean field results are so good, since the mean field method in $D=3$ dimensions is usually regarded as a rough approximation at best.  A possible answer is that in the effective Polyakov line actions each SU(3) spin is coupled to {\it many} other SU(3) spins on the lattice, and not merely to the nearest neighbors.\footnote{The strength of coupling to non-nearest neighbors falls roughly as $1/r^4$, up to the long-distance cutoff $r_{max}$.}  As a consequence, the basic idea behind mean field theory, i.e.\ that each spin is effectively coupled to the average spin on the lattice, may be a much better approximation to the true situation in the effective theories than one would suppose from prior experience with nearest-neighbor couplings.
   
   The second result is that in the case where the complex Langevin and mean field results differ, the difference occurs at chemical potentials where the Langevin method is clearly unreliable, due to the appearance of the M{\o}llgaard-Splittorff
branch cut problem \cite{Mollgaard:2013qra}.  A possible way around the branch cut difficulty is to complexify the SU(3) elements $U_\vx, U^\dg_\vx$, rather than the angles $\th_a(\vx)$, a strategy which is used for lattice gauge theory and which was already mentioned in \cite{Aarts:2011zn}.  In that case the exponentiation of the measure factor is avoided, and there is no branch cut problem.  On the other hand, one must still monitor the Langevin evolution of $U_\vx, U^\dg_\vx$ in the complex plane, to see if large violations of the unitarity contraint $U_\vx U^\dg_\vx = \mathbbm{1}$, caused by large excursions into the complex plane, are avoided.  This approach is a possible direction for future work.

    It is significant that the disagreement between the mean field and complex Langevin methods only arises at values of the chemical potential where the complex Langevin method fails.  Of course, a failure of complex Langevin does not imply that the corresponding mean field results are necessarily correct; it could be that both are wrong.  At the moment we have no independent check.  What {\it can} be said at this stage is that mean field theory applied to effective Polyakov line actions, where it has been checked against the reliable results of an alternate method, works remarkably well.  It is possible that, given the
effective Polyakov line action for a gauge-matter system obtained by relative weights, the theory can be solved without resorting to any further numerical simulation.
    
\acknowledgments{I thank Gert Aarts for a helpful discussion.  This research is supported in part by the
U.S.\ Department of Energy under Grant No.\ DE-FG03-92ER40711.   
   
\bibliography{pline}
 
\end{document}